\documentclass[review,12pt]{elsarticle}
\usepackage[utf8]{inputenc}
\usepackage[toc,page]{appendix}
\usepackage{amssymb}
\usepackage{babel}
\usepackage{graphicx}
\usepackage{booktabs}
\usepackage{xcolor}
\usepackage{gensymb} 
\usepackage{amsmath,amsfonts,amsthm,bm} 
\usepackage{mathrsfs} 
\usepackage[margin=2cm]{geometry}
\usepackage{gensymb}
\usepackage{multirow}
\usepackage[skip=6pt plus1pt, indent=40pt]{parskip}
\usepackage{setspace} 
\usepackage{appendix}
\usepackage{upgreek}

\journal{npj Materials Degradation}
\usepackage{graphicx}
\usepackage{caption}
\usepackage{subcaption}
\usepackage{float}
\usepackage{hyperref}
\hypersetup{colorlinks=true, linkcolor=blue, citecolor=black, filecolor=magenta, urlcolor=black}
\usepackage{soul}
\usepackage{accents}
\usepackage{color,soul} 
\usepackage{color} 
\usepackage{xcolor,soul}
\biboptions{sort&compress}
\sethlcolor{lightgray}

\begin{document}
\begin{frontmatter}
\title{A mesoscale phase-field model of intergranular liquid lithium corrosion of ferritic/martensitic steels}

\author[1] {Alexandre Lhoest}

\author[2] {Sasa Kovacevic}

\author[3] {Duc Nguyen-Manh}

\author[3] {Joven Lim}

\author[2,4] {Emilio Mart\'inez-Pa\~neda}

\author[1] {Mark R. Wenman}

\address[1]{Imperial College London, Centre for Nuclear Engineering, South Kensington Campus, London SW7 2AZ, UK}
\address[2]{Department of Engineering Science, University of Oxford, Oxford OX1 3PJ, UK}
\address[3]{United Kingdom Atomic Energy Authority, Culham Campus, Abingdon OX14 3DB, UK}
\address[4]{Corresponding author(s) (email address): emilio.martinez-paneda@eng.ox.ac.uk}

\begin{abstract}
A phase-field model is developed to simulate intergranular corrosion of ferritic/martensitic steels exposed to liquid lithium. The chromium concentration of the material is used to track the mass transport within the metal and liquid (corrosive) phase. The framework naturally captures intergranular corrosion by enhancing the diffusion of chromium along grain boundaries relative to the grain bulk with no special treatment for the corrosion front evolution. The formulation applies to arbitrary 2D and 3D polycrystalline geometries. The framework reproduces experimental measurements of weight loss and corrosion depth for a 9 wt\% Cr ferritic/martensitic steel exposed to static lithium at 600 $\degree$C. A sensitivity analysis, varying near-surface grain density, grain size, and chromium depletion thickness, highlights the microstructural influence in the corrosion process. Moreover, the significance of saturation is considered and evaluated. Simulation results show that near-surface grain density is a deciding factor, whereas grain size dictates the susceptibility to intergranular corrosion.\\
\end{abstract}


\end{frontmatter}

\section{Introduction} \label{intro}

The commercialization of fusion energy remains hindered by many overwhelming engineering problems that need addressing prior to its implementation. A key one is the feasibility of breeding tritium \textit{in situ} using lithium-bearing materials. Liquid lithium (Li) provides unique benefits \cite{Smith1981BlanketReactors, Shanliang2003NeutronicMaterials}. The concern of radiation damage is eliminated as the breeder material is in a liquid state, which offers greater flexibility and efficacy in the extraction of tritium as the breeder material can circulate away from the extreme conditions of the reactor. Furthermore, the inherent efficacious thermal properties of liquid metals indicate that one can operate at higher temperatures, leading to greater thermal efficiencies \cite{Malang1995ComparisonBlankets, Kirillov2006RFTests}. Several potential candidates for the optimum structural material for future fusion power plants are currently being investigated, including ferritic/martensitic (F/M) steels and vanadium-based alloys \cite{Bloom1998TheSystems, Giancarli2006BreedingDEMO, Smith1998MaterialsSystems}. Each candidate material possesses its respective advantages, yet owed to the worldwide expertise and generations of use in the fission industry, many of the proposed designs have opted for F/M steels \cite{Giancarli2020OverviewActivities}, thereby exploiting the high strength and low creep rate at elevated temperatures plus desirable irradiation performance \cite{Klueh2005ElevatedReactors}.

A limiting concern to the implementation of liquid Li as a breeder material is the harsh corrosive environment it produces when in contact with structural alloys, ultimately leading to weight loss and surface recession (i.e., wall thinning) influencing the longevity of the structural material and heightening the risk of accumulated activated material \cite{Chopra1984CompatibilityReactors}. The corrosion of alloys with liquid Li is solely based on physio-chemical processes, where the principal mechanisms are solubility-driven dissolution, intergranular corrosion (IGC), and mass transfer \cite{ Tortorelli1984LiquidDevelopment, Tortorelli1988CorrosionReactions}. As such, the composition of the structural alloy in contact with the liquid metal is a substantial factor in the corrosion resistance, owed to the varying solubility limits of alloying elements differing by several orders of magnitude in some cases \cite{Lyublinski1995NumericalEutectic}. Furthermore, Li preferentially attacks grain boundaries (GBs) due to their heterogeneity, tendency for chemical segregation, and precipitation of secondary phases. The compatibility of F/M steels with liquid Li is heavily correlated with the amount of Cr in the composition of the steel \cite{Chopra1988CompatibilitySystems}. Cr is preferentially leached from the structural alloy attributed to its comparatively heightened solubility limit \cite{Chopra1988CompatibilitySystems,Tortorelli1981CorrosionApplications}. The amount of Cr regulates the chemical activity of carbon in the metal by forming metallic carbides (i.e., Cr$_{23}$C$_6$). Liquid Li preferentially corrodes GBs as these regions are decorated with carbides, relative to a tempered martensitic microstructure \cite{Ruedl1982IntergranularSteels}. Once exposed to liquid Li, these carbide precipitates become unstable and are subsequently leached \cite{Bell1989TheLithium} via the following reaction
\begin{equation}
    \text{Cr}_{23} \text{C}_{6\,(\mathrm{s})}+6 \text{Li}_{(\mathrm{l})}\rightarrow 3 \text{Li}_2 \text{C}_{2\, (\mathrm{sol})}+23 \text{Cr}_{(\mathrm{sol})}
    \label{metal_dissolution}
\end{equation}
resulting in a heavily chemically altered region \cite{Hosaka2022ChemicalNaK}, commonly observed through a phase transformation of the matrix from martensite to ferrite \cite{Xu2007CompatibilityLithium}. Once the alloying elements in the immediate vicinity are depleted, resulting in a chemical gradient, bulk alloying elements diffuse towards the corroding interface to replenish the lost material. Although the compatibility of F/M steels with liquid Li at elevated temperatures has been experimentally explored in depth \cite{Tsisar2011StructuralCoarsening, Zhang2024StudyLithium}, showcasing its applicability, there still remains little consensus on the dominant corrosion mechanism and the principal microstructural features which govern the corrosion process.

Evaluating IGC resistance and the long-term effects of materials exposed to liquid Li presents significant challenges due to the unique conditions of this corrosive environment. Numerical modelling can expand the reach of experimental studies by providing insights and extending the duration of experiments involving liquid Li under various conditions. Ideally, these models can guide the design of materials optimized for liquid Li environments and suggest strategies to minimize susceptibility to IGC. Modelling the corrosion resistance of structural materials has been detailed in abundance through various computational techniques, including Lagrangian-Eulerian approaches \cite{Sarkar2012ADissolution, Sun2014AnCorrosion}, peridynamics \cite{Jafarzadeh2019PittingModels, Chen2015PeridynamicDamage}, cellular automata models \cite{Gong2022NucleationAnalysis, Fatoba2018SimulationApproach} and phase-field methods \cite{Martinez-Paneda2024Phase-fieldResearch, Cui2023Electro-chemo-mechanicalImplementation, Makuch2024ACracking, Ansari2019ModelingMetals, Lin2021Phase-fieldCracking, Tantratian2022PredictingModel, Brewick2022SimulatingModeling, Amador2024QuantitativePhenomena, Chadwick2018NumericalMethods, Kandekar2024MasteringScheme, Li2023NewCorrosion}. However, these are mainly used in the context of uniform and localized corrosion and stress corrosion cracking. As such, these models are limited in capturing IGC, assessing the compatibility of liquid Li with F/M steels, and incorporating the microstructural influence on IGC. The synergistic effect of aggressive environments and underlying microstructures has been recently resolved in the context of IGC \cite{Jafarzadeh2018PeridynamicDamage, Guiso2020IntergranularAutomata, Guiso2022IntergranularAutomata, Ansari2020Multi-Phase-FieldMaterials, VivekBhave2023AnSalt}. Albeit these IGC models provide mechanistic predictions and insight, they do not properly resolve the differing corrosion kinetics in the grain bulk and GBs \cite{Jafarzadeh2018PeridynamicDamage, Guiso2020IntergranularAutomata, Guiso2022IntergranularAutomata}, a key feature of liquid Li IGC. Other models utilize multiphase-field formulations \cite{Ansari2020Multi-Phase-FieldMaterials, VivekBhave2023AnSalt} to distinguish between GBs, grain interiors, and the corrosive environment. However, these multiphase-field approaches inherently possess high computational costs. Moreover, neither model surveyed the correlation between microstructural properties and their influence on the IGC process. The present work aims to develop a phase-field model to assess the IGC of F/M steels in contact with liquid Li. As the underlying physical mechanics in the corrosion process, the diffusion process is differentiated between GBs and grain bulk, effectively and naturally addressing the varying corrosion kinetics while keeping the phase-field equation unaltered. An additional stationary parameter is introduced to distinguish mass transport between GBs and grain interiors, which reduces the computational cost intrinsically associated with multiphase-field formulations.   

The remainder of the paper is organized as follows. The underlying corrosion mechanism of F/M steels exposed to liquid Li and the associated modelling assumptions are presented in the following section. The chromium (Cr) concentration of the material is used to track mass transport within the metal and liquid (corrosive) phase. The phase-field model is subsequently derived from a generalized thermodynamic free energy functional. The governing equations for the phase-field parameter and Cr concentration in the system, as the primary variable of the model, are derived in a thermodynamically consistent way. GBs are differentiated from grain interiors using an additional independent stationary parameter that allows for incorporating representative diffusivities of Cr along GBs and grain bulk, exploiting physical material parameters to capture IGC. The constructed framework is calibrated and validated against experimental measurements on a 9 wt\% Cr F/M steel in contact with static liquid Li at 600 $\degree$C in Section \ref{setup}, demonstrating its ability to capture IGC phenomena. After validation, a sensitivity analysis for different
microstructural features, including grain density at the exposed surface, grain size, and Cr depletion thickness, is performed to investigate their role in the corrosion resistance of the simulated material in Section \ref{results}. The main conclusions are subsequently discussed in Section \ref{Dis} as well as the investigation findings and recommendations for future work.

\section{Results} \label{results}

\subsection{Influence of grain density at the exposed surface} \label{GB_results}

\begin{figure}[h!]
    \centering
    \includegraphics[width=16cm]{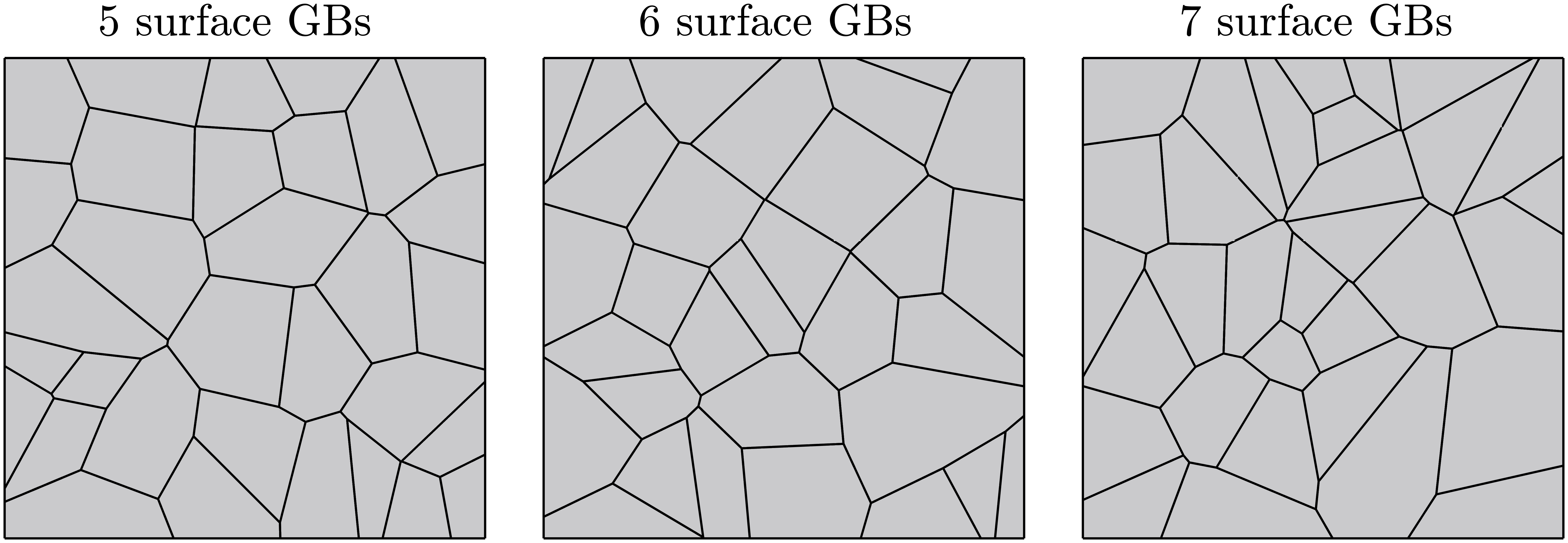}
    \captionsetup{labelfont = bf,justification = raggedright}
    \caption{\textbf{Representative microstructures for the 5, 6 and 7 surface GB microstructure simulated.} Each microstructure has an average grain size of 20 $\upmu$m, while varying the number of GBs at the exposed surface located at the upper most boundary.}
    \label{GB_rep_microstructure}
\end{figure}

\begin{figure}[h!]
    \centering
    \includegraphics[width=16cm]{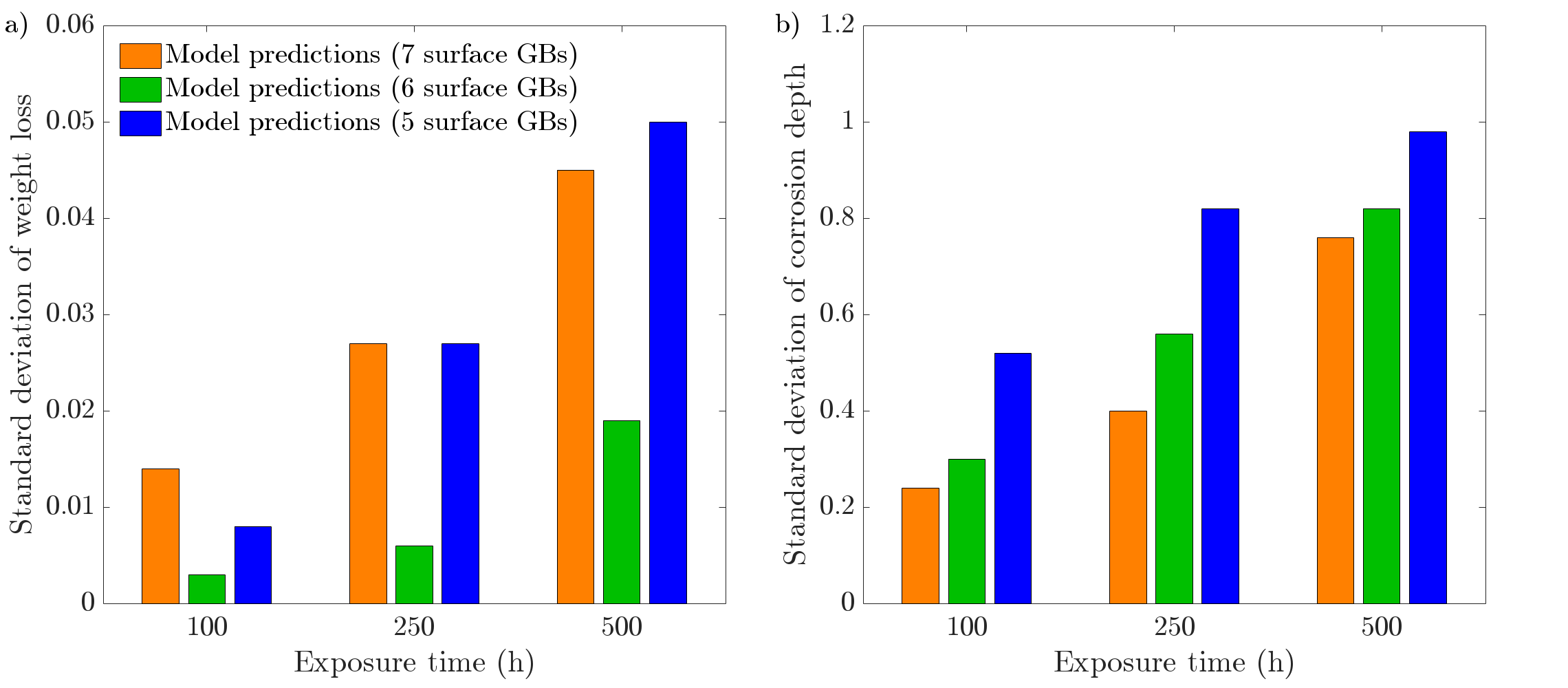}
    \captionsetup{labelfont = bf,justification = raggedright}
    \caption{\textbf{Standard deviations at 100, 250, and 500 hours intervals for 20 $\upmu$m average grain size microstructure with 5, 6, and 7 GBs present at the exposed surface.} Standard deviations for \textbf{a} weight loss and \textbf{b} corrosion depth. The data presented for each case is taken from an average of ten microstructures.}
    \label{GB_WL_CD_SD}
\end{figure}

The microstructures employed to calibrate the model had 6 GBs at the exposed surface. To understand the influence of grain density at the near-surface, the grain structure distribution is varied while keeping an average grain size of 20 $\upmu$m. In addition to the reference geometry in Section \ref{setup}, two further microstructures, with 5 and 7 GBs at the exposed surface, were analysed. As previously, the average behaviour was taken from ten equiaxed microstructures for each case. Fig. \ref{GB_rep_microstructure} displays the representative microstructure for each scenario while the ten microstructures used for the 5 GB and 7 GB microstructures can be viewed in Figs. S.2 and S.3 (Supplementary Information). The resultant phase-field predictions can be seen in Fig. \ref{GB_WL_CD} where the standard deviation (SD) data at 100, 250, and 500 hours exposure time are depicted in Fig. \ref{GB_WL_CD_SD}.

The weight loss data highlights a positive correlation with the number of grains at the exposed surface. The microstructures that possess 7 GBs exhibited the most significant weight loss over the 500-hour simulation. Contrastingly, the microstructures with 5 GBs give rise to the lowest weight loss. Overall, all three microstructures with 5, 6, and 7 GBs at the exposed surface produce largely identical profiles, whereby the difference in weight loss between the three types of microstructures increases with increasing exposure time. As such, it is apparent during the first 500 hours of exposure that the rate of weight loss is most severe for the 7 GB microstructure, resulting in a comparatively greater weight loss of 1.34 g/m$^2$. The 5 GB microstructure, on the other hand, exhibited 1.01 g/m$^2$ of material weight loss. On average, the total weight loss after 500 hours increases by $\sim$15\% each time, increasing the grain density at the interface by a single grain. Consequently, the 7 GB microstructures overestimated the weight loss at 100 and 250 hours compared to the experimental measurements (Section \ref{setup}). Yet, it remains within the relative error for the 250-hour weight loss experimental measurement. Alternatively, the 5 GB microstructures predicted with a degree of high accuracy the weight loss at 100 hours. However, it underestimates the weight loss at 250 hours, while remaining within the relative error. Furthermore, after 500 hours, the 5 and 7 GB microstructures give rise to almost identical SDs, which are more than twice as large compared to the 6 GB microstructures, Fig. \ref{GB_WL_CD_SD}(a). The corrosion depth data displays zero influence on the near-surface grain density as all three types of grain structures give rise to predominantly identical corrosion depth profiles. Overall, the three microstructures simulated generate an average corrosion depth of 12.3 $\upmu$m. Contrary to the weight loss data, the SDs regarding the corrosion depth exhibit a clear order. The 5 GB microstructures produce the greatest SD, followed by the 6 GB microstructures, where the 7 GB microstructures return the smallest SD after 500 hours of exposure time, Fig. \ref{GB_WL_CD_SD}(b).

\begin{figure}[h!]
    \centering
    \includegraphics[width=16cm]{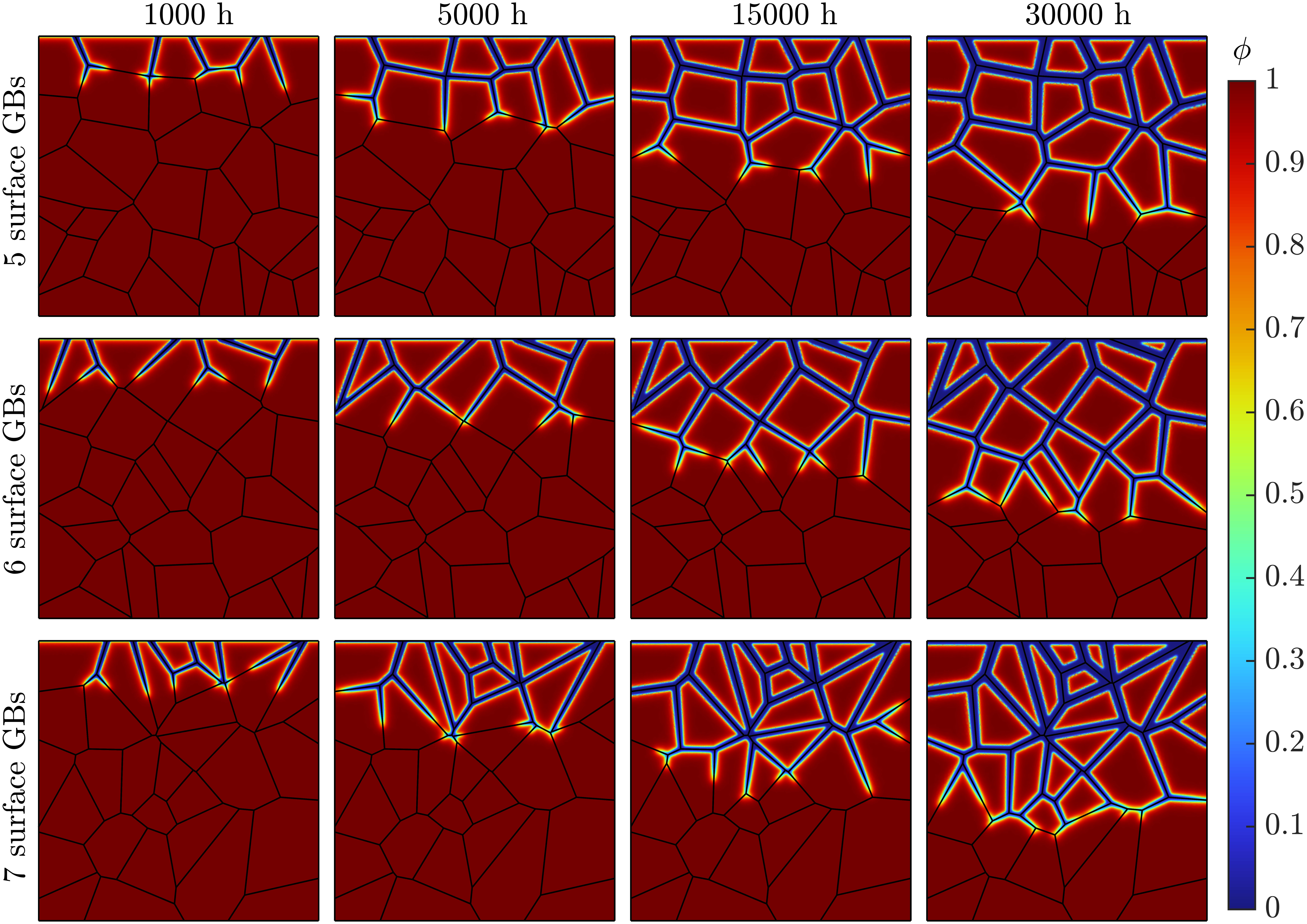}
    \captionsetup{labelfont = bf,justification = raggedright}
    \caption{\textbf{Long-term IGC evolution while varying the near-surface grain density.} The phase-field variable $\phi$ in the simulated metallic specimen during a 30,000-hour simulation for the three representative microstructures possessing 5, 6, and 7 GBs at the exposed surface.}
    \label{GB_IGC_phi}
\end{figure}

\begin{figure}[h!]
    \centering
    \includegraphics[width=16cm]{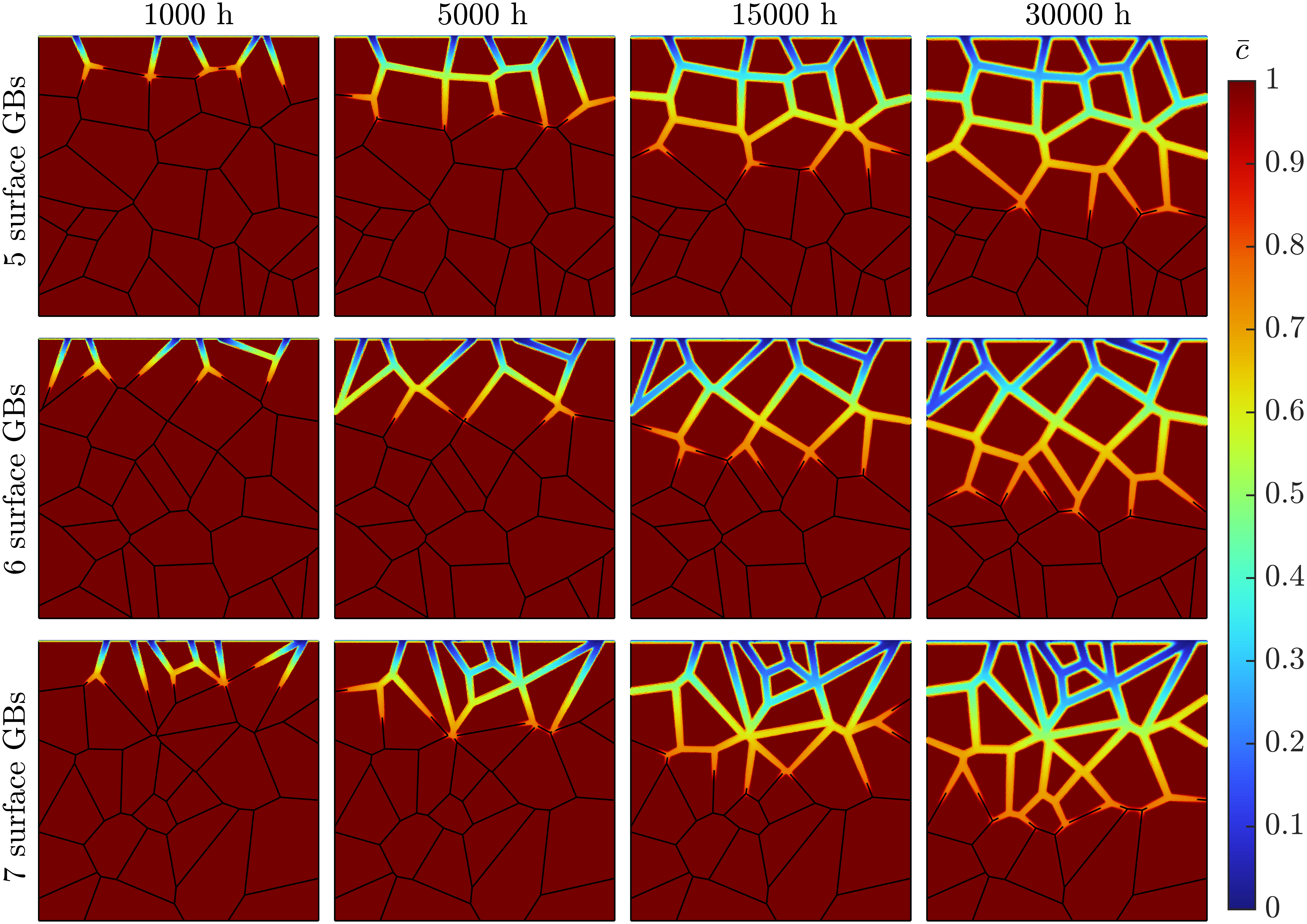}
    \captionsetup{labelfont = bf,justification = raggedright}
    \caption{\textbf{Long-term intergranular leaching of Cr while varying the near-surface grain density.} The normalized concentration of Cr $\overline{c}$ in the simulated metallic specimen during a 30,000-hour simulation for the three representative microstructures possessing 5, 6, and 7 GBs at the exposed surface.}
    \label{GB_IGC_conc}
\end{figure}

Fig. \ref{GB_IGC_phi} shows the progression of corrosion of the representative microstructures (Fig. \ref{GB_rep_microstructure}) following a 30,000-hour exposure time. The onset of corrosion initiates at the GBs exposed to the liquid metal at the upper most boundary. As time increases, the evolution of $\phi$ continues along the GBs deeper into the material. Congruently, Fig.\ref{GB_IGC_conc} shows the diminished concentration of Cr intensifies as Li advances into the steel. The spread of concentration of Cr along the GBs indicates that the corrosion process operates in the diffusion-controlled regime. The near-surface GBs are seen to be completely leached of Cr, with this process increasing in depth as time progresses. In addition, the degree of corrosion regarding the near-surface GBs visibly heightens, as seen through the increased thickness of the $\phi=0$ region. In some instances, two corrosion fronts are seen advancing at either end of the same GB, where they ultimately meet, engulfing the entire grain. The expansive nature of Li to diffuse across all GBs is evident, as it proceeds deeper into the material, clearly leaving behind a deteriorated material. After 30,000 hours of exposure time, the liquid Li corrodes most GBs in the simulated solid phase. However, it only causes marginal corrosion to the grains on the exposed surface. 

\subsection{Influence of grain size} \label{GS_results}

\begin{figure}[h!]
    \centering
    \includegraphics[width=16cm]{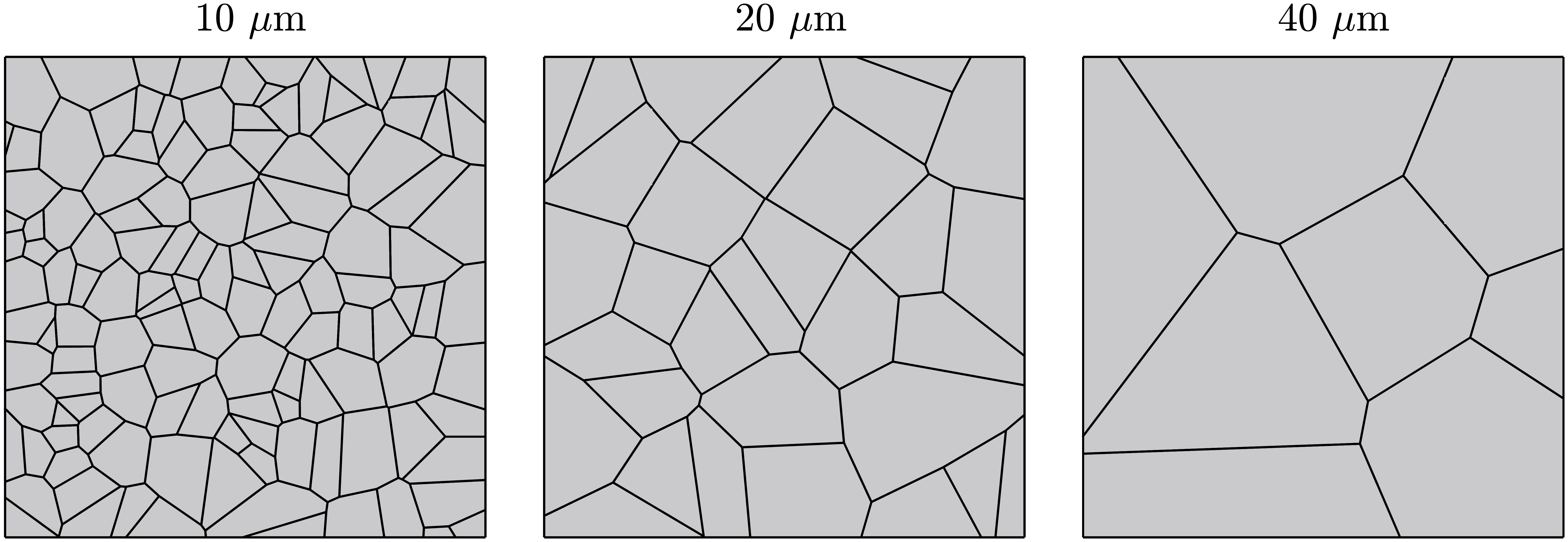}
    \captionsetup{labelfont = bf,justification = raggedright}
    \caption{\textbf{Representative microstructures for the three average grain sizes simulated.} The GBs present at the exposed surface are 10, 6, and 2 for the 10, 20, and 40 $\upmu$m microstructures.}
    \label{GS_rep_microstructure}
\end{figure}

\begin{figure}[h!]
    \centering
    \includegraphics[width=16cm]{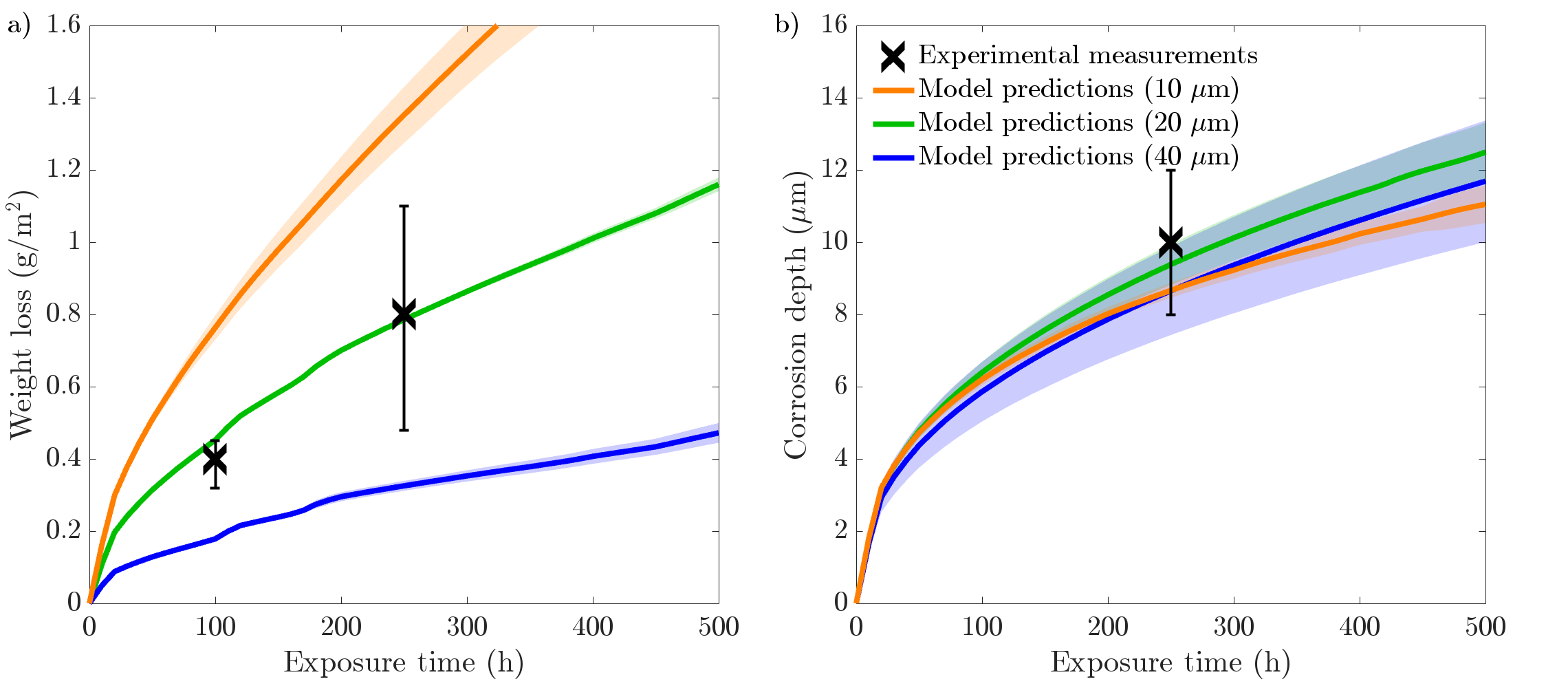}
    \captionsetup{labelfont = bf,justification = raggedright}
    \caption{\textbf{Phase-field predictions following a 500-hour simulation for 10, 20, and 40 $\upmu$m average grain size microstructures compared to experimental measurements \cite{Xu2008CorrosionConditions}.} Phase-field predictions for \textbf{a} weight loss and \textbf{b} corrosion depth. The data presented for each case is taken from an average of ten microstructures where the shaded coloured region represents the SD of the ten simulations.}
    \label{GS_WL_CD}
\end{figure}

\begin{figure}[h!]
    \centering
    \includegraphics[width=16cm]{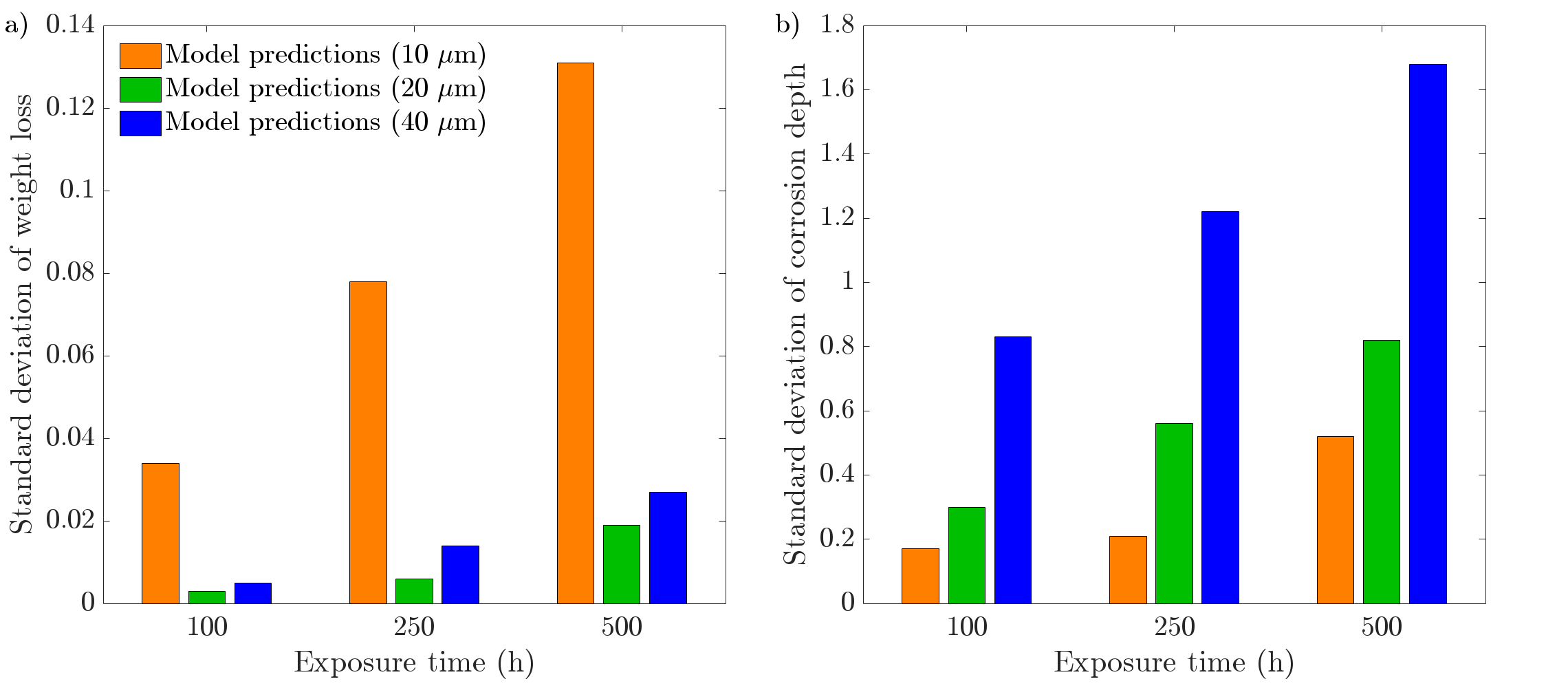}
    \captionsetup{labelfont = bf,justification = raggedright}
    \caption{\textbf{Standard deviations at 100, 250, and 500 hours intervals for 10, 20, and 40 $\upmu$m average grain size microstructures.} Standard deviations for \textbf{a} weight loss and \textbf{b} corrosion depth. The data presented for each case is taken from an average of ten microstructures.}
    \label{GS_WL_CD_SD}
\end{figure}

The grain size plays a governing role in IGC due to the synergistic relationship with GB density within the bulk of the material. As such, the average grain size is altered to observe this dependency. Two additional average grain sizes of 10 and 40 $\upmu$m are modelled. To isolate the effect of grain size from near-surface grain density, all three average grain size microstructures should possess the same number of GBs at the exposed surface. Nonetheless, the feasibility of a 10 and 40 $\upmu$m equiaxed microstructures possessing the same number of grains at the exposed surface is unattainable. Therefore, although the near-surface grain density may differ between the three grain sizes (10, 20, and 40 $\upmu$m), they remain constant within the ten microstructures for each case. As such, the 10 and 40 $\upmu$m microstructures possess 10 and 2 GBs at the exposed surface. Representative microstructures for all three grain sizes are shown in Fig. \ref{GS_rep_microstructure}, while the entire set of microstructures for the 10 and 40 $\upmu$m microstructures can be seen in Figs. S.4 and S.5 (Supplementary Information). The resultant weight loss and corrosion depth following a 500-hour simulation are given in Fig. \ref{GS_WL_CD}. The SD data at 100, 250, and 500 hours exposure time can be seen in Fig. \ref{GS_WL_CD_SD}.

The weight loss shown in Fig. \ref{GS_WL_CD}(a) showcases that the 10 $\upmu$m average grain size microstructures exhibits the most severe weight loss after 500 hours of exposure time. Alternatively, the largest grain size simulated of 40 $\upmu$m displays the highest corrosion resistance. Unsurprisingly, neither the 10 $\upmu$m nor the 40 $\upmu$m microstructures coincide with the experimental data extracted from a 20 $\upmu$m average grain size F/M specimen. Fig. \ref{GS_WL_CD}(b) portrays that the average corrosion depth reached by liquid Li is largely independent of the grain size. Although there is little difference between the corrosion depths, there is nonetheless a noticeable pattern in the SDs for each grain size simulated, whereby the 10 $\upmu$m grain size possesses the smallest SDs and the 40 $\upmu$m microstructures give rise to the largest SD, more than twice as large as the SDs generated from the 20 $\upmu$m microstructures, after 500 hours (Fig. \ref{GS_WL_CD_SD}(b)). The SDs regarding the weight loss show an exceeding variation for the 10 $\upmu$m microstructures, whereas the 20 $\upmu$m and 40 $\upmu$m produce similar minute variances. 

\begin{figure}[h!]
    \centering
    \includegraphics[width=16cm]{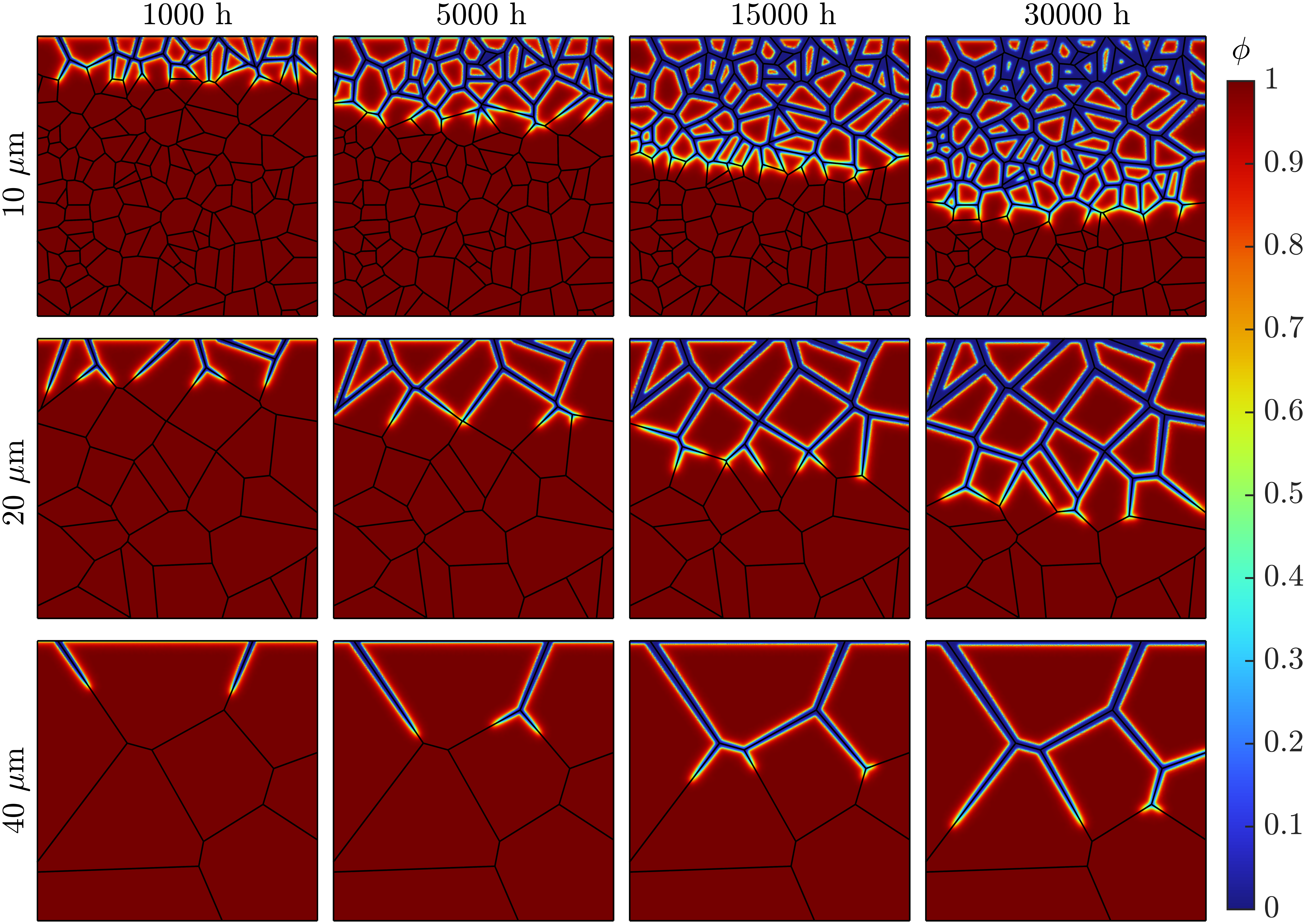}
    \captionsetup{labelfont = bf,justification = raggedright}
    \caption{\textbf{Long-term IGC evolution while varying the grain size.} The phase-field variable $\phi$ in the simulated metallic specimen during a 30,000-hour simulation for the three representative microstructures with average grain sizes of 10, 20, and 40 $\upmu$m.}
    \label{GS_IGC_phi}
\end{figure}

\begin{figure}[h!]
    \centering
    \includegraphics[width=16cm]{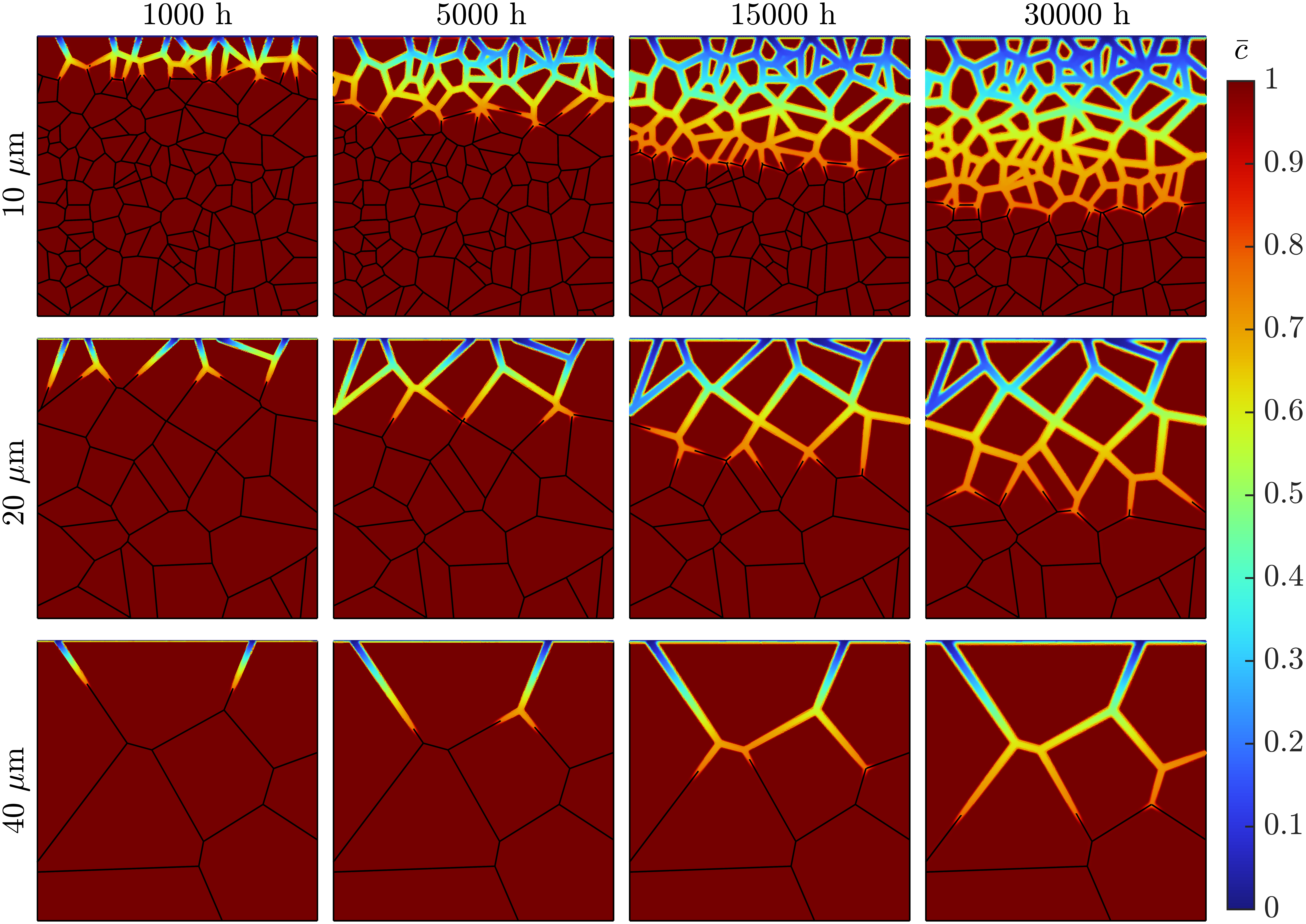}
    \captionsetup{labelfont = bf,justification = raggedright}
    \caption{\textbf{Long-term intergranular leaching of Cr while varying the grain size.} The normalized concentration of Cr $\overline{c}$ in the simulated metallic specimen during a 30,000-hour simulation for the three representative microstructures with average grain sizes of 10, 20, and 40 $\upmu$m.}
    \label{GS_IGC_conc}
\end{figure}

Fig. \ref{GS_IGC_phi} highlight the IGC evolution across the representative microstructures (Fig. \ref{GS_rep_microstructure}) for 30,000 hours of exposure time. Analogous to varying the near-surface grain density, the detrimental impact of Li corrosion is apparent, given by the extensive reach during the exposure time. The correlation between the average grain size and severity of corrosion is similarly made clear. The 40 $\upmu$m microstructure experiences considerably less degradation via the reduced presence of Li within the material. The low density of GBs, resulting in comparatively long GBs with little GB branching, limited the deterioration following exposure to liquid Li. Alternatively, the 10 $\upmu$m microstructure suffered extreme IGC over the exposure time, established through the vast area of the simulated domain whereby $\phi=0$. Furthermore, due to the relatively finer grain size, grains are more easily engulfed by Li containing GBs, leaving behind an intrigue network of corroded GBs with vastly diminished concentrations of Cr, Fig. \ref{GS_IGC_conc}.

\subsection{Influence of the thickness of the smeared GB region} \label{lp_results}

As detailed in Section \ref{PFM}, the computational GB thickness $l_p$ dictates the thickness of the smeared GB region that has increased diffusivity relative to the grains. In addition, implementation of the physical thickness of Cr depletion at the GB, $\delta_{gb}$, circumvents the issue with modelling nanometer-sized features far from their physical dimensions and thus avoids computationally expensive simulations. As such, $\delta_{gb}$ is kept constant while $l_p$ is altered to analyse its impact on the corrosion process of the simulated specimen. For this case, the representative microstructures for the 5, 6, and 7 GB microstructures (Fig. \ref{GB_rep_microstructure}) were selected. Only a single run is completed for each change in $l_p$, as running all ten microstructures for each case is not needed to establish a correlation. Fig. \ref{lp_WL_CD} shows the weight loss and corrosion depth for the three microstructures and the values of $l_p$ chosen: 50, 100, and 200 nm.

For all three types of grain structures with 5, 6, and 7 GBs at the exposed surface, the 50 nm thickness gives rise to the lowest weight loss and 200 nm thickness results in the most extensive weight loss. That said, the number of GB entry points remains a deciding factor in weight loss. This is most clearly seen for the 7 GB microstructure with 100 nm thickness results in more significant weight loss than for the 5 GB microstructure with 200 nm thickness. This is additionally valid for the 7 GB microstructure with 50 nm thickness when compared against the 5 GB microstructure with 100 nm thickness. On average between all three microstructures simulated, decreasing the computational GB thickness from 100 nm to 50 nm, results in a final weight loss drop of 19.7$\%$ whereas when increased from 100 nm to 200 nm the final weight loss increases by 24.7$\%$. As presented in Fig. \ref{GB_WL_CD}, there is no obvious relationship between the corrosion depth and the near-surface grain density. Consequentially, varying $l_p$ produces three distinctive groups as opposed to the weight loss data, which is clearly spread. It is clear that the microstructures with 50 nm thickness suffer the greatest corrosion depth, and those with 200 nm thickness have the highest resistance to Li penetration.  

\subsection{Influence of saturation} \label{liquid_results}

As stated in Section \ref{setup}, a concentration sink is implemented to encapsulate the leaching effect of liquid Li in static conditions. This, however, omits the impact of saturation, which hinders and ultimately halts the corrosion process, leading the corrosion data to plateau \cite{Xu2008CorrosionConditions}. To prove the capability of the current model to capture the effect of saturation and to observe whether a concentration sink might influence the corrosion data, a small liquid phase in contact with the solid phase is introduced whereby the concentration sink boundary condition is removed. The incorporation of a liquid phase necessitates a liquid phase diffusivity. That said, as the IGC process depends on the solid-state diffusivity, the value of the liquid phase diffusivity is not of significant importance as long as it is greater than the GB diffusivity. The liquid phase diffusivity of Cr in static liquid Li has been reported to be $2.2\times10^{-8}$ m$^2$/s at 570 $\degree$C \cite{Griaznov1989InteractionConditions}. However, to keep the diffusion coefficient gradient between phases at a manageable difference to avoid unnecessary computational cost, the liquid phase diffusivity is assigned to be $D_{\mathrm{gb}}$ as this will always be greater than $D^{\prime}_{\mathrm{gb}}$.

\begin{figure}[h!]
    \centering
    \includegraphics[width=8cm]{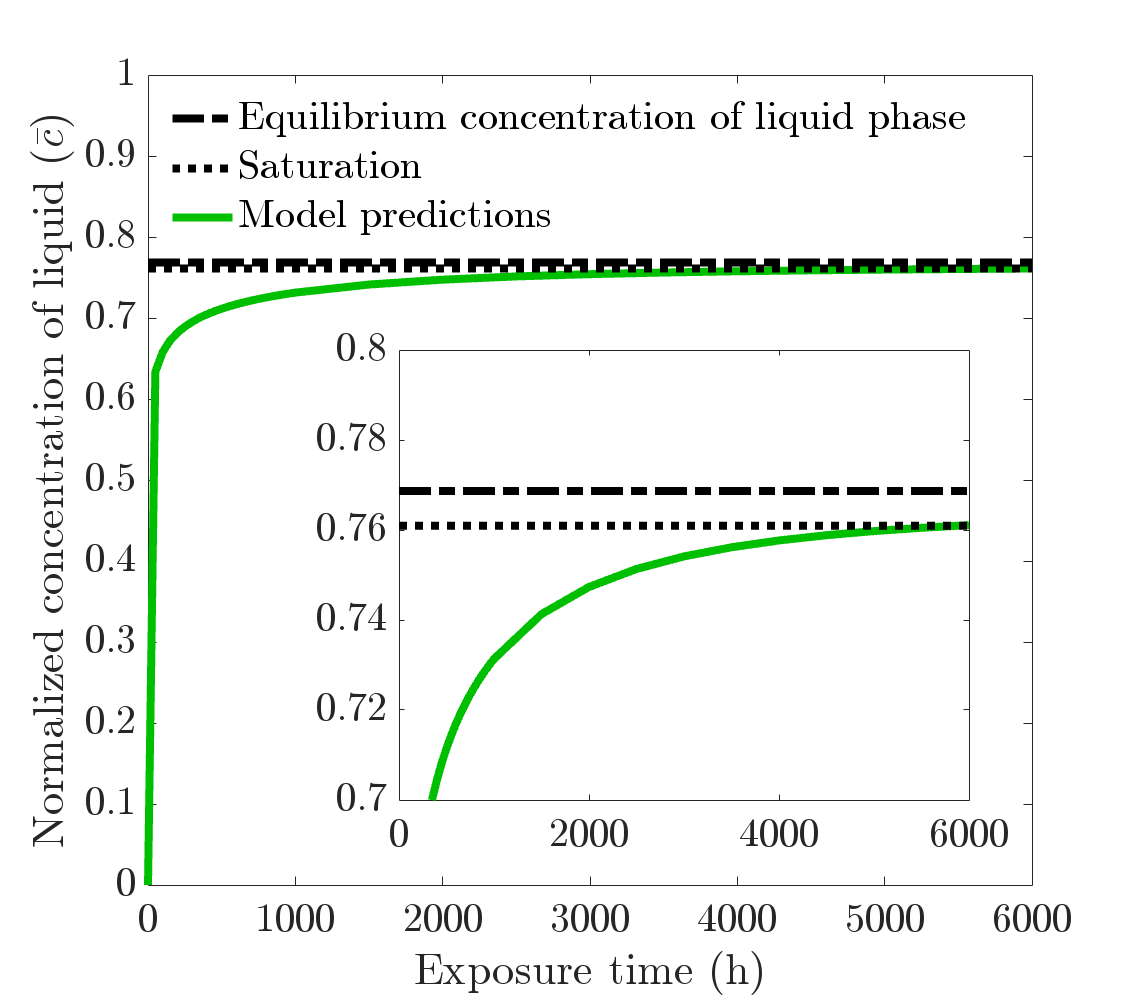}
    \captionsetup{labelfont = bf,justification = raggedright}
    \caption{\textbf{Normalized concentration of Cr in the liquid phase as a function of exposure time.} With a 1 $\upmu$m thick liquid phase above a 20 $\upmu$m average grain size microstructure with 6 GBs, it takes 6000 hours for the system to reach saturation. The data presented here is an average from ten microstructures where the shaded coloured region represents the SD of the ten simulations. Note, the consistency across the ten microstructures resulted in a negligible SD. Sub-graph displays a magnified version to clearly highlight the liquid concentration nearing saturation. All subsequent concentration plots possess a range and domain akin to the sub-graph.}
    \label{liquid_saturation}
\end{figure}

As the normalized concentration of Cr in the liquid phase ($\overline{c}$) approaches the normalized equilibrium concentration of the liquid phase ($\overline{c}_{\mathrm{l,eq}}$), the chemical free energy density decreases thereby reducing the chemical driving force of the corrosion process, Eqs. (\ref{eq5}) and (\ref{fchem}). The diminished chemical driving force decreases the rate of dissolution, thereby producing an asymptotic relationship between $\overline{c}$ and $\overline{c}_{\mathrm{l,eq}}$. As such, herein saturation of the system is considered met when $\overline{c}\ge99\%\cdot \overline{c}_{\mathrm{l,eq}}$. It is consequentially found when simulating a 1 $\upmu$m thick liquid phase in contact with the reference microstructure from Section \ref{setup} (i.e., 20 $\upmu$m average grain size with 6 GBs at the exposed surface) that saturation is achieved after 6000 hours. This is shown in Fig. \ref{liquid_saturation}, where the normalized concentration of the liquid phase quickly approaches and eventually reaches the condition of saturation. The thickness of 1 $\upmu$m is chosen to reach saturation relatively quickly. Consequentially, all subsequent simulations involving the liquid phase are set with a 6000-hour exposure time with a 1 $\upmu$m thick liquid phase. To survey the influence of the liquid phase, and thus, the effect of saturation on the corrosion process, the liquid phase is introduced to the models described in Section \ref{GB_results} and Section \ref{GS_results}. 

\begin{figure}[h!]
    \centering
    \includegraphics[width=16cm]{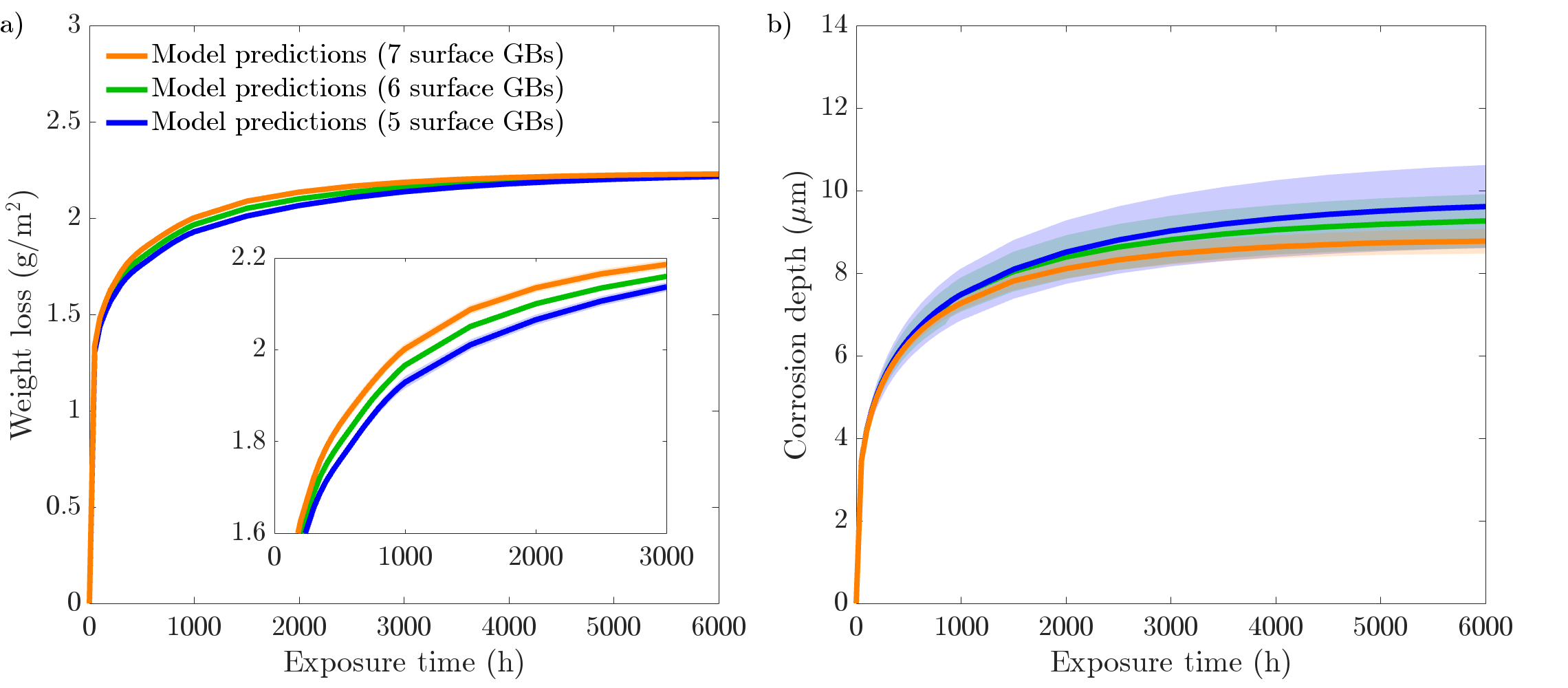}
    \captionsetup{labelfont = bf,justification = raggedright}
    \caption{\textbf{Phase-field predictions following a 6000-hour simulation for 20 $\upmu$m average grain size microstructure with 5, 6, and 7 GBs present at the exposed surface with a 1 $\upmu$m liquid phase in contact with the solid phase.} Phase-field predictions for \textbf{a} weight loss and \textbf{b} corrosion depth. The data presented for each case is taken from an average of ten microstructures each where the shaded coloured regions represents the SD of the ten simulations. Sub-graph displays a magnified version to clearly distinguish between the weight loss profiles.}
    \label{Liquid_GB_WL_CD}
\end{figure}

\begin{figure}[h!]
    \centering
    \includegraphics[width=16cm]{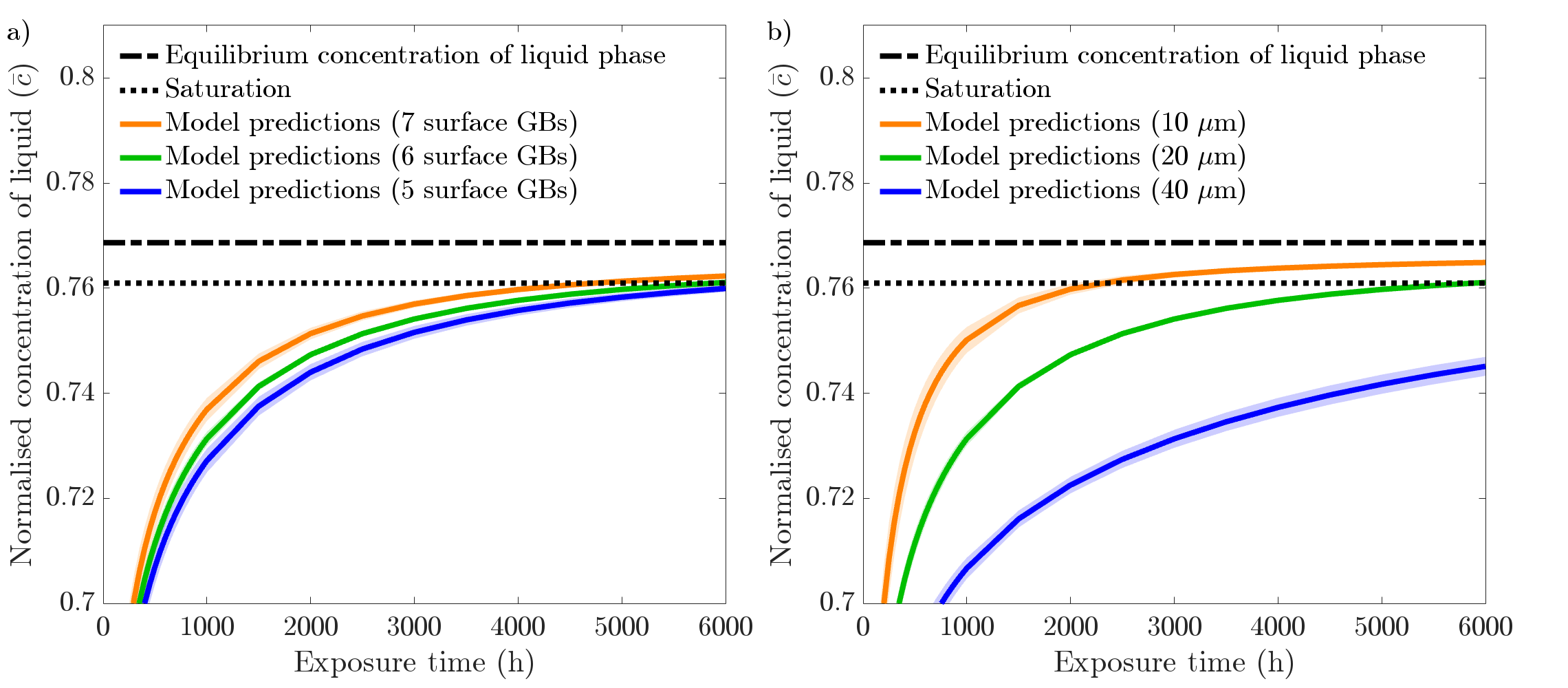}
    \captionsetup{labelfont = bf,justification = raggedright}
    \caption{\textbf{Normalized concentration of Cr in the liquid phase as a function of exposure time while varying microstructural properties.} Normalized concentration of Cr in the liquid while varying \textbf{a} the near-surface grain density for a 20 $\upmu$m average grain size microstructure and \textbf{b} the average grain size. The data presented for each case is an average of ten microstructures where the shaded coloured regions represents the SD of the ten simulations.}
    \label{Liquid_GB+GS_Conc}
\end{figure}

The corrosion data for varying the near-surface grain density can be seen in Fig. \ref{Liquid_GB_WL_CD}. As all the microstructures with 5, 6, and 7 GBs are in contact with an identical quantity of liquid phase, the amount of dissolved Cr needed to reach saturation is constant, thereby producing, if not identical weight loss profiles resulting in largely consistent final weight loss of 2.22 g/m$^2$ following 6000 hours exposure time, Fig. \ref{Liquid_GB_WL_CD}(a). Nonetheless, the weight loss experienced by the 7 GB microstructures plateaus before the other two grain structures, indicating the system reached saturation quicker. Therefore, the corrosion process of leaching halted earlier. Alternatively, the 5 GB microstructures take noticeably longer for the weight loss to eventually plateau and terminate. Fig. \ref{Liquid_GB+GS_Conc}(a) reinforces this notion, with the 7 GB microstructures clearly reaching saturation within a shorter exposure time than that of the 6 GB, followed by the 5 GB microstructures. Interestingly, contrary to the results displayed in Section \ref{GB_results} where a concentration sink is implemented, the SDs for the weight loss when saturation is considered are firstly too narrow to have any influential significance and secondly are all consistent across the three types of microstructural conditions simulated. The effect of saturation on the corrosion process is also seen in Fig. \ref{Liquid_GB_WL_CD}(b), where corrosion depth eventually plateaus. In this case, there is, although marginal, a clear trend emerging. The 5 GB microstructures produce the deepest depth, followed by the 6 GB, and finally, the 7 GB microstructures give rise to the shallowest corrosion depth. The SDs of the corrosion depths remain in the same order as that found with the concentration sink, where the 5 GB microstructures possess the greatest followed by 6 GB and finally 7 GB microstructures.

\begin{figure}[h!] 
    \centering
    \includegraphics[width=16cm]{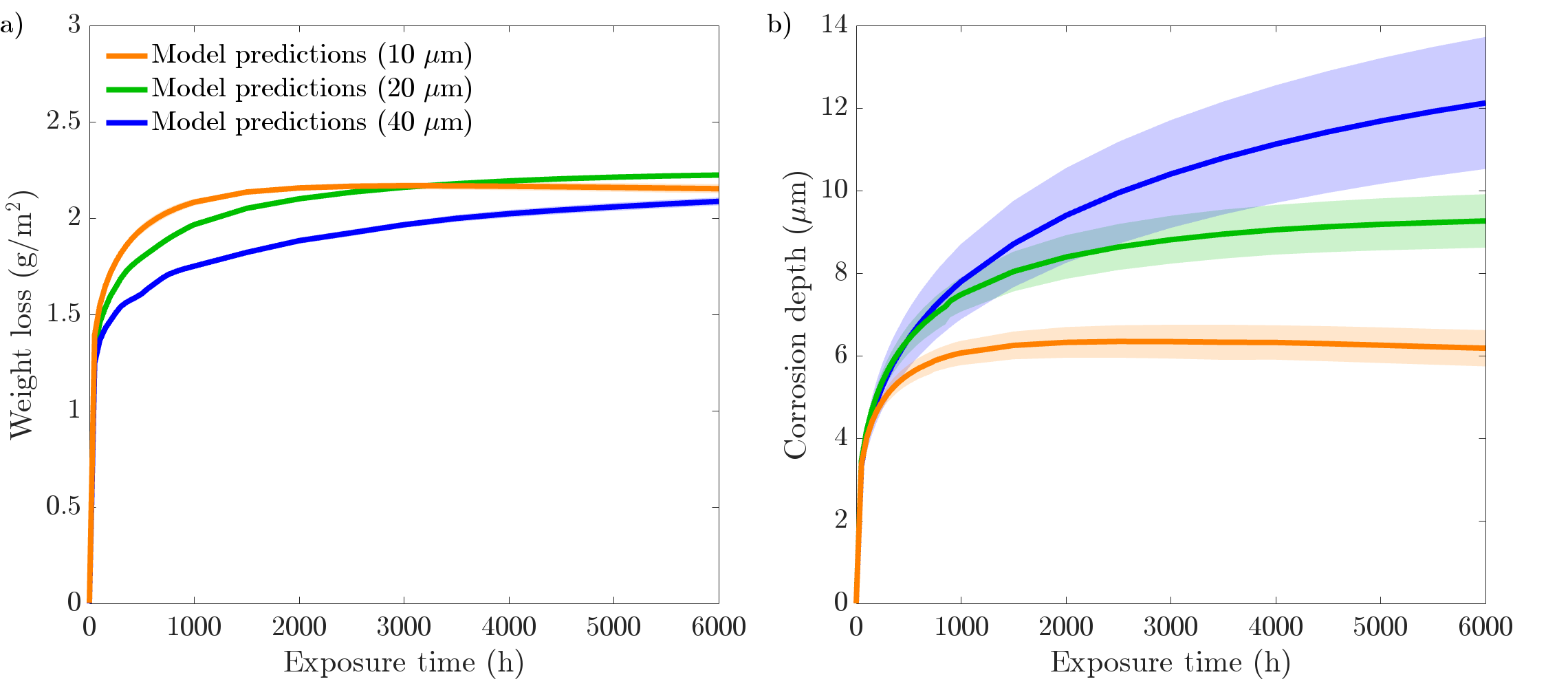}
    \captionsetup{labelfont = bf,justification = raggedright}
    \caption{\textbf{Phase-field predictions following a 6000-hour simulation for 10, 20, and 40 $\upmu$m average grain size with a 1 $\upmu$m liquid phase in contact with the solid phase.} Phase-field predictions for \textbf{a} weight loss and \textbf{b} corrosion depth. The data presented for each case is taken from an average of ten microstructures each where the shaded coloured regions represents the SD of the ten simulations.}
    \label{Liquid_GS_WL_CD}
\end{figure}

Additional analysis is carried out in which the 1 $\upmu$m thick liquid phase is implemented while varying the average grain size from 10, 20, and 40 $\upmu$m. Initially, the weight loss data, displayed in Fig. \ref{Liquid_GS_WL_CD}(a), shows that the grain size and rate of dissolution are positively correlated. Moreover, the 10 $\upmu$m microstructures reaches saturation first, indicative of the comparatively early plateau of the weight loss profile. With respect to the 40 $\upmu$m microstructures, the weight loss profile has yet to plateau, indicating the system did not reach saturation. This is confirmed by Fig. \ref{Liquid_GB+GS_Conc}(b) whereby the finer grain-sized microstructures bring the system to saturation at the fastest rate. In addition, it is further evident from Fig. \ref{Liquid_GB+GS_Conc}(b) that the 40 $\upmu$m microstructures did not reach saturation within the 6000 hours exposure time. Interestingly, however, is the progression of the weight loss profile for the 10 $\upmu$m microstructures, as it peaks around 3000 hours, where thereafter begins to dip and drop below the weight loss profile for the 20 $\upmu$m microstructures. As the same quantity of liquid phase is implemented throughout, the total weight loss experienced strongly resembles the final weight loss when varying the near-surface grain density. Another similarity between Fig. \ref{Liquid_GB_WL_CD}(a) and Fig. \ref{Liquid_GS_WL_CD}(a) is the negligible SDs. Fig. \ref{Liquid_GS_WL_CD}(b) displays a clear pattern in the corrosion depth when altering the average grain size of the microstructure. As depicted, the 40 $\upmu$m microstructures show the deepest corrosion depth, whereas the 10 $\upmu$m microstructures experiences the shallowest corrosion depth. Furthermore, similar to the weight loss data regarding the 10 $\upmu$m microstructures, the corrosion depth peaks and subsequently recedes. Analogous to when the concentration sink is implemented, the SDs for the corrosion depth remain in the same order starting at 10, 20, and 40 $\upmu$m in increasing order. 

\section{Discussion} \label{Dis}
\begin{figure}[h!]
    \centering
    \includegraphics[width=16cm]{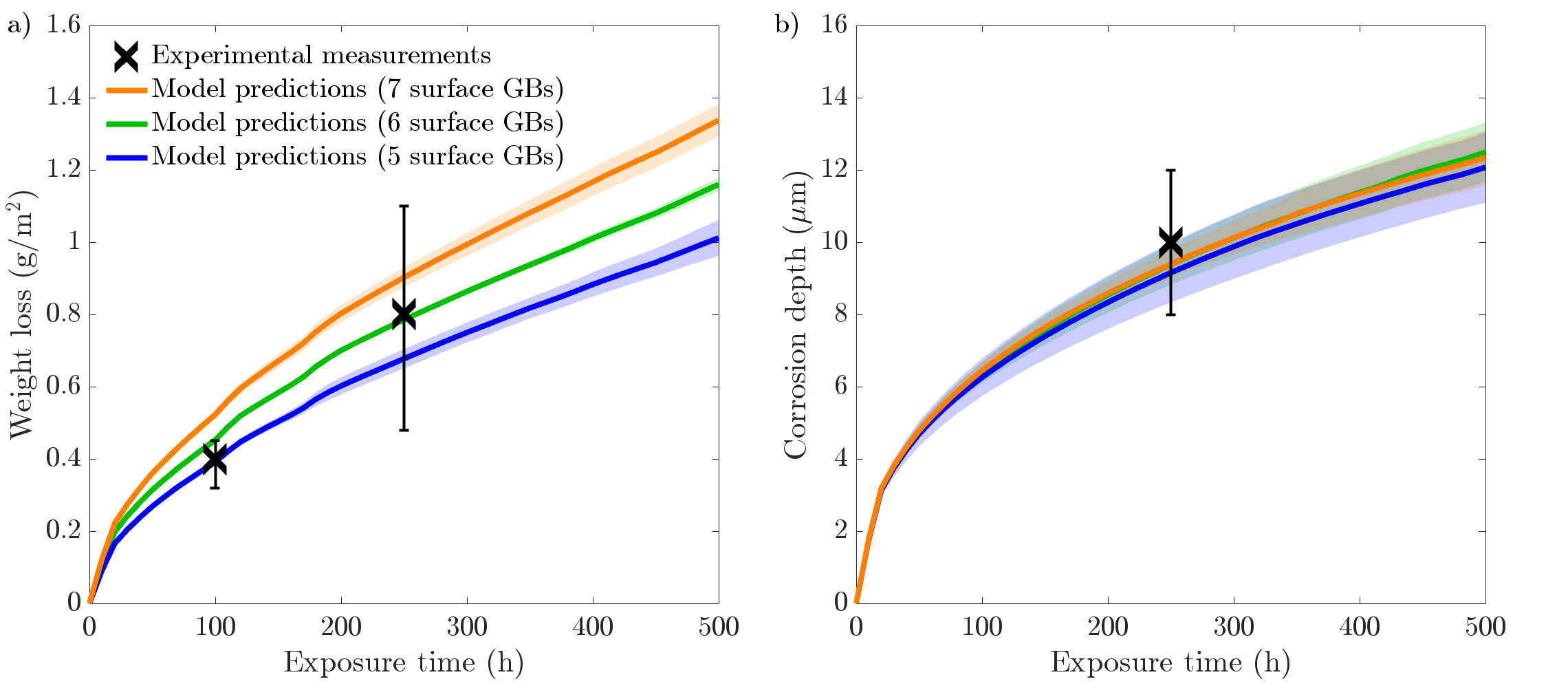}
    \captionsetup{labelfont = bf,justification = raggedright}
    \caption{\textbf{Phase-field predictions following a 500-hour simulation for 20 $\upmu$m average grain size microstructure with 5, 6, and 7 GBs present at the exposed surface compared to experimental measurements \cite{Xu2008CorrosionConditions}.} Phase-field predictions for \textbf{a} weight loss and \textbf{b} corrosion depth. The data presented for each case is taken from an average of ten microstructures where the shaded coloured regions represents the SD of the ten simulations.}
    \label{GB_WL_CD}
\end{figure}

It is clear that the most susceptible microstructures to weight loss possess the greatest number of near-surface grains giving rise to more GB entry points to infiltrate into the material, Fig. \ref{GB_WL_CD}(a). To emphasize this trend, the inverse projected grain size ($G$), where $G$ is the ratio of the exposed surface length against the number of near-surface grains, was plotted against the total weight loss (not shown here for brevity). This yielded an $R^2$ of 0.9973 closely aligning with Bhave \textit{et al.} \cite{VivekBhave2023AnSalt} reported $R^2$ of 0.9966, consolidating the significance of GB entry points in IGC. The initial weight loss is dictated by the number of GB entry points, where thereafter, the total length of GBs available to corrode, governed by the grain size, offers a more comprehensive indicator to the degree of dissolution. Reducing the average grain size dramatically increases the density of GBs and thus amplifies the weight loss. Furthermore, the greater variation in the distribution of GBs among the 10 $\upmu$m microstructures, impacting the amount of accessible GBs, resulted in high variances in the weight loss, Fig. \ref{GS_WL_CD_SD}(a), thus behaving more inconsistently relative to the larger grain sizes. The inverse relationship between grain size and IGC susceptibility has been similarly observed experimentally \cite{Tortorelli1984MassLithium, Li2013Corrosion873K}, highlighting the capabilities of the model.   

As the corrosion process can continue indefinitely, via the concentration sink, the relationship between corrosion depth and microstructural features is not evident, Figs. \ref{GB_WL_CD}(b) and \ref{GS_WL_CD}(b). Nonetheless, it can offer insight into governing traits that dictate the susceptibility of structural materials to corrosion depth. The large variance in corrosion depths for the 40 $\upmu$m microstructures suggests that penetration depth is highly sensitive to the microstructure when the average grain size is large, Fig. \ref{GS_WL_CD_SD}(b). The relative orientation of GBs greatly effects the ability of Li to penetrate into the material, whereby the significance of their orientation increases with fewer GB entry points. Furthermore, the degree of GB branching governs the length Li must diffuse before changing its course, placing additional significance on their orientation as it strictly controls the depth Li can reach. Therefore, larger grain sizes produce a greater spread in corrosion depths as opposed to finer grain sizes. This suggest, in light of predictability, that finer grained structural materials are more desirable to optimize consistency in relation to IGC. That said, a holistic approach need consider the increased susceptibility of IGC as a consequence of the reduced grain size in order to determine the prime microstructure \cite{Hariharan2024MicrostructureDesign}. Likewise when varying the near-surface grain density; the 7 GB microstructures initially possess, albeit marginally, more entry points through which Li penetrates and, therefore, the relative orientation of the GBs has less influence on the depth of corrosion resulting in a lower variance, Fig. \ref{GB_WL_CD_SD}(b).

Replacement of the concentration sink with the effect of saturation in the corrosion process shifts microstructural influence on the corrosion behaviour. Due to the limitless corrosion process brought by the concentration sink, the relationship between the microstructural features (i.e., near-surface grain density and grain size) and weight loss is evident whereas the corrosion depth is independent of any microstructural changes. On the other hand, incorporating a minor quantity of liquid phase, thereby implementing saturation, clearly highlights the synergy between corrosion depth and microstructural features, yet consequentially dampens the correlation with weight loss. It should be noted, the conditions brought by the concentration sink strongly mimic the conditions of a dynamic breeder loop. The thermal gradients of dynamic loops produce detrimental corrosive conditions created by the perpetual dissolution and deposition of material; mass transfer, thus inhibiting saturation \cite{Tortorelli1988CorrosionReactions, Chopra1984CompatibilityReactors}. Consequentially, evaluating the corrosion behaviour between static (i.e. saturation) and dynamic (i.e. concentration sink) conditions offers insight into the varied compatibility of F/M steels with liquid Li when transitioning from a controlled environment to higher fidelity conditions.

It is apparent that the near-surface grain density dictates the rate of saturation, as it offers more entry points to initiate penetration. Therefore, a greater number of GB entry points yields a faster rate of dissolution thus reaching saturation quicker, Fig. \ref{Liquid_GB+GS_Conc}(a). This, in turn, controls the corrosion duration before the process is halted, ultimately dictating the depth Li is able to penetrate. The average grain size, via the bulk GB density, has greater influence on the rate of saturation, impacting more heavily the corrosion depth. Moreover, the 40 $\upmu$m microstructures had yet to reach saturation as seen through the sustained severity of the corrosion depth, Figs. \ref{Liquid_GB+GS_Conc}(b) and \ref{Liquid_GS_WL_CD}(b). Regarding the variance in the corrosion behaviour, the introduction of saturation does not affect the order of corrosion depth SDs remaining as 7 GBs $<$ 6 GB $<$ 5 GB and 10 $\upmu$m $<$ 20 $\upmu$m $<$ 40 $\upmu$m. Nonetheless, the SDs on weight loss are significantly less prominent when saturation is accounted for. Overall, high density of near-surface grains coupled with small grain size maximizes the rate of saturation, thereby reducing the corrosion duration, and in turn, producing shallower corrosion depths, aligning with experimental reports \cite{DiStefano1964CorrosionLithium}. It is additionally observed that the corrosion depth and the weight loss for the 10 $\upmu$m microstructures peak where thereafter it decreases, Fig. \ref{Liquid_GS_WL_CD}. As mentioned, near saturation the chemical free energy density decreases considerably, and for the 10 $\upmu$m microstructures, it becomes smaller than the interfacial free energy term, thus making the latter the dominant term in Eq. (\ref{eqnAC}). Consequentially, the phase-field interface becomes more rigid and thus becomes inapt in simulating complex morphologies causing the IGC pathways to become narrower and shorter during the later stages of exposure, resulting in an apparent diminished weight loss and corrosion depth.

Figs. \ref{GB_IGC_phi} and \ref{GS_IGC_phi} illustrates the IGC evolution with respect to $\phi$, showcasing the diffusive behaviour of Li through the material. As the exposure time progresses, near-surface grains are engulfed by corroded GBs, yet the grains remain largely uncorroded. This becomes more apparent as the average grain size decreases. In reality, as corroded GBs surround a grain, they become susceptible to detachment from the bulk matrix, as observed experimentally \cite{Kondo2010Erosion-corrosionImpeller, Kondo2011FlowBreeders}, producing a surface pebbled morphology. Although the corrosion depth was independent of microstructural features, it is overwhelming apparent that the degree of Li infiltration is strongly correlated to the near-surface and bulk GB density. The greater density of GBs additionally exposes more material to leaching, exacerbating the extent of Cr dissolution in the material, Figs. \ref{GB_IGC_conc} and \ref{GS_IGC_conc}. 

\begin{figure}[h!]
    \centering
    \includegraphics[width=16cm]{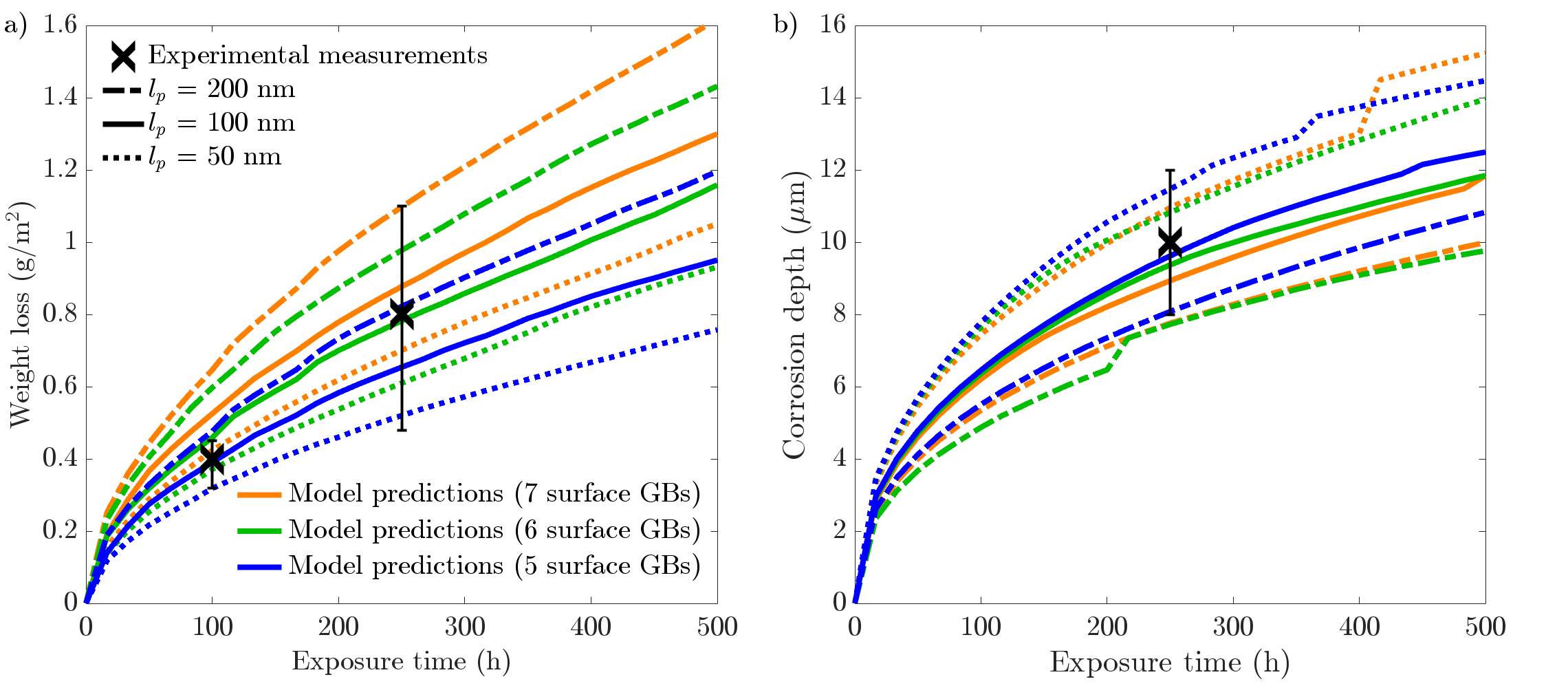}
    \captionsetup{labelfont = bf,justification = raggedright}
    \caption{\textbf{Phase-field predictions following a 500-hour simulation for 20 $\upmu$m average grain size geometry with 5, 6, and 7 GBs present at the exposed surface while varying $l_p$ by 50, 100 and 200 nm compared to experimental measurements \cite{Xu2008CorrosionConditions}.} Phase-field predictions for \textbf{a} weight loss and \textbf{b} corrosion depth. The coloured lines in the legend denote the microstructure (i.e., number of near-surface GBs) whereas the black lines of varying line-type illustrate the computational GB thickness $l_p$. In conjunction they signify the conditions of each dataset.}
    \label{lp_WL_CD}
\end{figure}

The value of $l_p$ governs the thickness of the Cr depletion region, thus dictating the carbide-rich region whereby Li can penetrate and leach out alloying elements influencing the weight loss. Naturally, a positive correlation arises between $l_p$ and weight loss, Fig. \ref{lp_WL_CD}(a). Additionally, attributed to the constant product approach, the value of $l_p$ directly affects the GB diffusivity ($D^{\prime}_{\mathrm{gb}}$) ultimately altering the kinetics of the reaction. Consequentially, in increasing $l_p$, Li is unable to penetrate as deep into the material leading to a decreased average corrosion depth, Fig. \ref{lp_WL_CD}(b). It should be stressed that the width of GB Cr depletion should not impact the kinetics of Li penetration; the observed correlation is purely artificial and attributed to the constant product approach. Nonetheless, the influence on weight loss is valid, which emphasizes the importance of regulating the degree of alloy segregation at the GBs to optimize the resistance of high Cr F/M steels to liquid Li. Moreover, slight kinks, resulting in a sudden jump in corrosion depth, are observed in Fig. \ref{lp_WL_CD}(b).

Albeit displaying key microstructural features to improve the IGC susceptibility of steel alloys, there remain limitations that dampen the fidelity of the model. To align with the experimental work of Xu \textit{et al.} \cite{Xu2008CorrosionConditions}, a computationally intensive liquid phase model would be required, necessitating a concentration sink instead. Although a valid assumption prior to saturation, it permits a limitless corrosion process ultimately resulting in unrealistic behaviours over longer exposure times. Furthermore, the addition of a 1 $\upmu$m liquid phase required 6000 h for the reference microstructure to saturate, Fig. \ref{liquid_saturation}, conflicting with experimental data. The chemical free energy density curvature parameter $A$ is selected based on the accuracy of the phase-field predictions to the experimental data. This proportionality constant should be determined and applied to liquid Li systems to improve the capability of the current framework. The microstructures employed are not fully representative of a tempered martensite microstructure due to the absence of GB angles, thus omitting low-angle GBs (i.e., lath and block boundaries) which should be included to fully capture the microstructural influence of IGC. In addition, the model precluded the nucleation of corrosion products, namely the ternary nitride complex, a prevalent mechanism in liquid Li corrosion with steels \cite{Tsisar2011EffectLithium}. Its formation goes beyond perturbing the chemical potential of the system, thus exacerbating the corrosion process, yet additionally is thought to amplify the kinetics of penetration by fostering a favourable diffusive environment for Li \cite{Beskorovaynyi1983MechanismsLithium}.

Regarding the long-term performance of the material following exposure with liquid Li, confidence can be placed on the concentration sink model as this best resembles the continual dissolution conditions of a dynamic breeder blanket. Plus, the long-term behaviour of steels with respect to static systems is evidently dependent on the volume of liquid metal used. As detailed, the continual and indefinite corrosion process brought by the concentration sink results in largely identical corrosion depths independent of the grain structure of the specimen. As such, an average corrosion depth from the results displayed in Section \ref{GB_results} and Section \ref{GS_results}, following 30,000 hours, equating to over 3 years of exposure, results in 65.8 $\upmu$m. This extensive penetration into the material would inevitably cause severe degradation of the mechanical integrity of the material thereby necessitating routine replacement of the material within this time period. Additionally, the current model assumes the primary degradation mechanism to be intergranular penetration of Li via solubility-driven dissolution. Although this is a governing process in the compatibility between liquid Li and F/M steels, there remains other phenomena not included in the presented model which would likely exacerbate the penetration rate and thus need be considered to assess the longevity of these steels. A few notable examples include the impact of non-metallic impurities thus forming corrosion products, the influence of fluid flow and the effect of dissimilar metallic systems. Compounding these processes in a comprehensive model to accurately evaluate the performance of structural materials with liquid metals is paramount in identifying the ideal candidate for future fusion reactors.

The clear relationship between the microstructure of the specimen and its susceptibility to intergranular penetration detailed through the model predictions grants perspective on the key traits to alter, thereby maximizing the performance of structural alloys in contact with liquid Li. Under an indefinite dissolution process, the grain size should be enlarged, considering desired mechanical properties, to limit the amount of available GBs to diffuse through and thus corrode. Greater attention should be placed on expanding the size of PAGs as the corresponding GBs, due to their high angle nature, foster greater density of Cr$_{23}$C$_6$ type carbides. Reports have detailed that higher austenitization temperatures over a longer duration facilitate greater growth of austenite grains \cite{Lee2005EffectSteel,Souza2020AustenitizingMicroalloyed-Steel}. Another potential mitigation strategy is to limit the degree of Cr segregation to the GBs either by reducing the carbon content in the steel or by employing high-affinity carbide-forming elements such as vanadium, niobium, or tantalum \cite{Tortorelli1981CompatibilityLithium,Bell1989TheLithium}. This would additionally diminish the presence of free carbon within the matrix of the steels, reducing the tendency for non-metallic impurity interaction with liquid Li.

\begin{figure}[h!] 
    \centering
    \includegraphics[width = 15 cm]{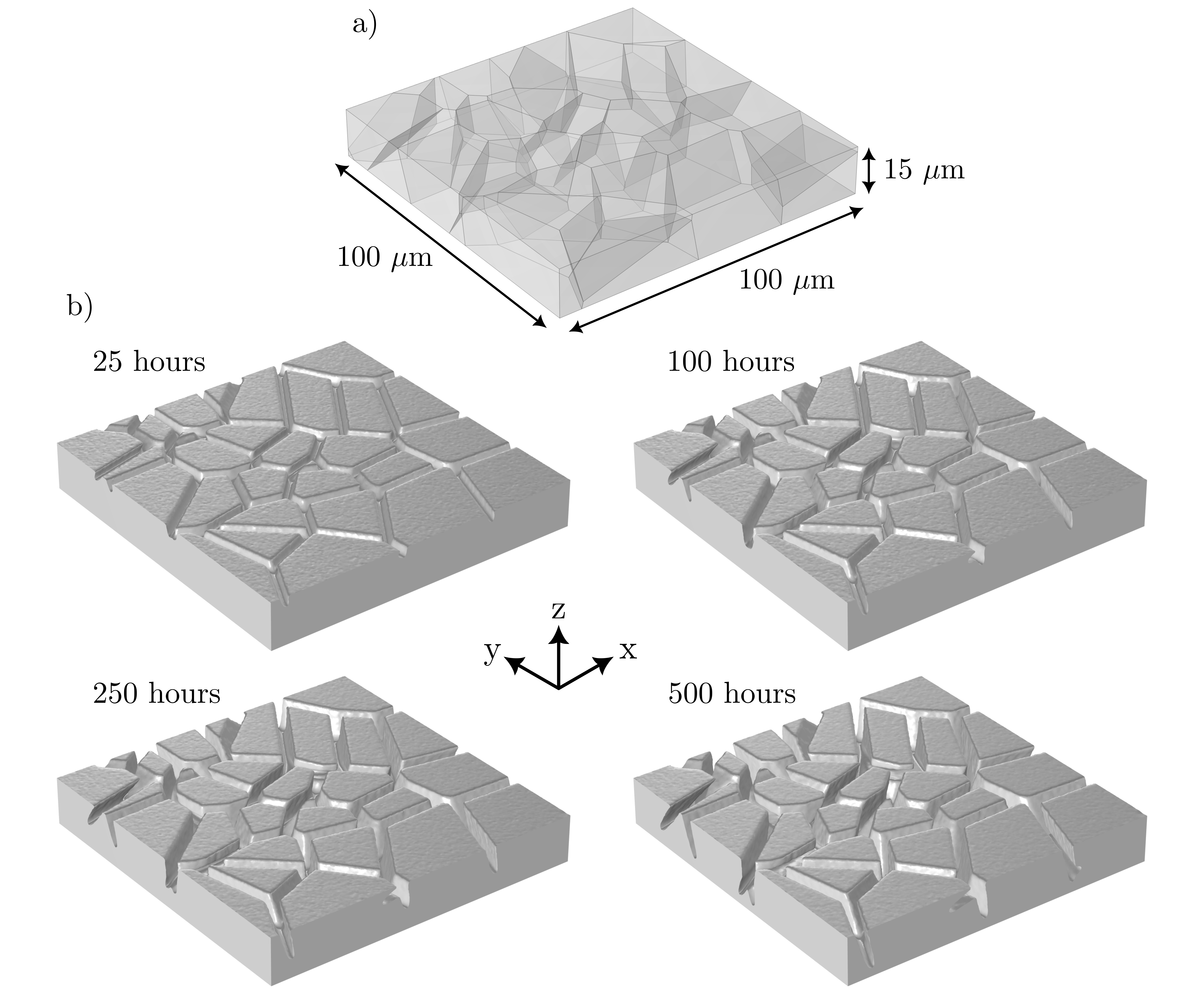}
    \captionsetup{labelfont = bf,justification = raggedright}
    \caption{\textbf{3D simulation of IGC.} Depiction of the \textbf{a} geometrical dimensions of the 3D model and \textbf{b} the intergranular corrosion evolution at 25, 100, 250 and 500 hours exposure time. Only the solid phase $\phi \ge 1/2$ is shown for clarity.}
    \label{3D}
\end{figure}

To demonstrate the capability of the current model as well as compare the implication in modelling varying dimensions, the results for a 3D simulation following 500 hours exposure time are shown in Fig. \ref{3D}. The geometry and computational conditions of the 3D model mimicked that of the reference 2D model detailed in Section \ref{setup}, i.e., 100 $\upmu$m in length and width with a depth of 15 $\upmu$m. The average grain size of the microstructure was 20 $\upmu$m with a concentration sink implemented on the upper-most boundary positioned at $z$ = 15 $\upmu$m. The depth of the material was selected to minimize the computational expense and was guided through the final corrosion depth reached following 500 hours exposure time with a concentration sink from the 2D models equating to $\sim$12 $\upmu$m, Fig. \ref{GB_WL_CD} and Fig. \ref{GS_WL_CD}. The evident evolution of the order parameter $\phi$ along the GBs can be seen in Fig. \ref{3D}. In accordance to the 2D model, the enhanced corrosion is isolated solely at the boundaries that possesses the heightened diffusivity, enabling the rapid intergranular extraction of alloying elements. The final weight loss of the 3D microstructure following 500 hours exposure time equated in 80.93 g/m$^2$, far exceeding the 1.16 g/m$^2$ experienced for the 2D microstructures detailed in Section \ref{setup}. It is important to emphasize that in the current formulation the interface kinetics coefficient $L$ is tailored to 2D microstructures. Therefore, $L$ ought to be derived relative to material properties to generate a more direct linkage between model predictions and experimental data. Bhave \textit{et al.} \cite{VivekBhave2023AnSalt} similarly reported greater mass loss from the 3D models relative to the 2D models in predicting the IGC behaviour of Ni-Cr alloys exposed to molten Li salt. Other similar studies \cite{Guiso2020IntergranularAutomata, Guiso2022IntergranularAutomata, Ansari2020Multi-Phase-FieldMaterials, Zhang2015SimulatingStructure, Nguyen2017Multi-phase-fieldMaterials, Simonovski2011ComputationalCracking} have been primarily focused on proof-of-concept of their respective formulations rather than analysing the significance regarding the corrosion data between 2D and 3D simulations. Nonetheless, 2D models have shown to yield representative data, that supports experimental theories, thus formulating our understanding of the microstructural influence on IGC. Future work should consider the role of the flow of the liquid phase, the nucleation of corrosion products from non-metallic impurities, the crystallographic orientation of grains, and GB angle (high and low-angle GBs), in mass transport and IGC of polycrystalline steels. It would additionally be warranted to conduct a parametric study consisting of a larger spectrum of microstructural dimensions (i.e., near-surface grain density and grain size) thus extracting more comprehensive data allowing deeper analysis of their impact on the IGC process.

In conclusion, a numerical framework based on the phase-field method is developed for assessing the IGC of polycrystalline steels with corrosive media. Herein, the model was applied to capture the leaching effects of liquid Li when in contact with a 9 wt\% Cr F/M steel. The key findings can be summarized as follows:
\begin{enumerate}
    \item The governing microstructural features that dominate and ultimately dictate the severity of corrosion are the number of GB entry points, i.e., near-surface grain density. The grain boundary density in the bulk, i.e., grain size, governs the susceptibility to Li corrosion. 
    \item The thickness of Cr depletion along the GBs $l_p$ plays a deciding role in the weight loss of the simulated material. This emphasizes the importance of limiting alloying GB segregation and consequential depletion in these regions, ideally hindering the ability for Li to diffuse intergranularly. 
    \item The effect of saturation is considered and compared against the use of a concentration sink. The former highlights the dependence of the corrosion depth on the microstructural features, whereas the latter amplifies the relationship with weight loss. Although the 10 $\upmu$m microstructures exhibit the fastest weight loss, this in turn results in the shallowest corrosion depth. With the idea of using grain engineering to maximize the compatibility with Li, it is worth dictating which corrosion behaviour is more detrimental to the structural integrity of the alloy.  
    \item The SDs regarding the corrosion depths shed light on the importance of the orientation of the GBs as well as their total length (i.e., GB density) in facilitating the penetration of Li into the material. It is shown that the 40 $\upmu$m microstructures yield varying corrosion depths as a result.
\end{enumerate}

\section{Methods}

\subsection{Phase-field model of intergranular corrosion} \label{PFM}

\begin{figure} [h!]
    \centering
    \includegraphics[width = 16 cm]{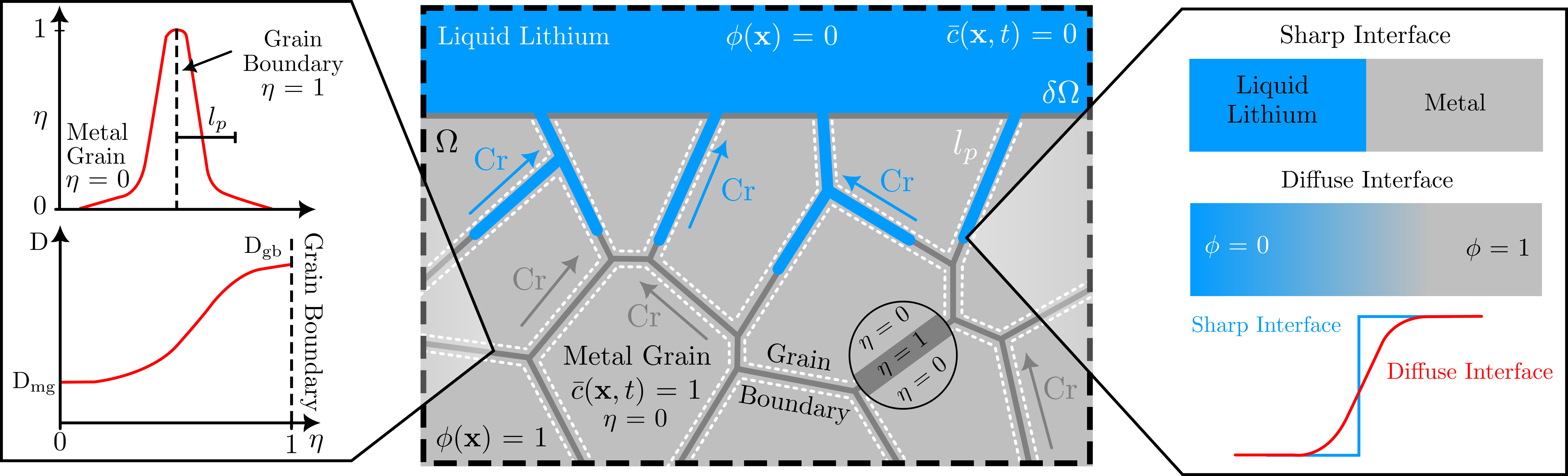}
    \captionsetup{labelfont = bf,justification = raggedright}
    \caption{\textbf{Polycrystalline material in contact with a corrosive environment highlighting the diffuse interface between the liquid ($\phi=0$) and solid ($\phi=1$) phases.} The GBs possess a heightened diffusivity $D_{\mathrm{gb}}$ compared to the metal grain $D_{\mathrm{mg}}$ via an additional parameter that differentiates between grain boundary ($\eta=1$) and metal grain ($\eta=0$). The corrosion mechanism is based on the bulk diffusion of Cr towards the exposed surface.}
    \label{pf_description}
\end{figure}

It is assumed in the current work the primary degradation mechanism is driven by the bulk diffusion of Cr in the metal phases. This assumptions aligns with \cite{Tsisar2011EffectLithium,Reeves1976GrainSteel,Patterson1975LithiumSteel}. The impact of other phases present in the metal (i.e., precipitates and inter-metallic species) and the effect of the Li$_2$C$_2$ by-product (Eq. \ref{metal_dissolution}) on the corrosion process is not considered in this study. The following kinematic variables are introduced to characterize the two phases. A continuous phase-field variable $\phi (\mathbf{x},t)$ is implemented to track each phase and the evolution of the corroding interface. $\phi=1$ represents the solid phase (i.e., F/M steel), $\phi=0$ defines the liquid corrosive agent (i.e., liquid Li), and $0<\phi<1$ describes the thin diffuse interface separating the opposing phases, Fig. \ref{pf_description}. As the composition of Cr in the F/M steels plays arguably the most central role in the corrosion process, modelling the entire composition of the steel is unwarranted. As such, the current model adopts a simplified F/M steel in the form of a simple binary Fe-Cr alloy. The metal phase in the present model is assumed to be Fe-9 wt\% Cr steel. It is further assumed in the present investigation that the material has a uniform composition and is composed of equiaxed grains separated by GBs. The composition of the material and its evolution during corrosion is characterized by the normalized concentration of Cr $\overline{c}(\mathbf{x},t)$ = $c/c_\mathrm{solid}$, where $c$ is the absolute concentration of Cr and $c_\mathrm{solid}$ the concentration of Cr initially present in the material. Two independent diffusion coefficients between the GB ($D_{\mathrm{gb}}$) and metal grains ($D_{\mathrm{mg}}$) are introduced to enhance the corrosion process along GBs, Fig. \ref{pf_description} whereby the employed values are displayed Table \ref{parameters}. The GB interpolation is achieved via an additional stationary parameter $\eta (\mathbf{x})$, that takes value $\eta=1$ at the GBs and $\eta=0$ elsewhere, as discussed below in Section \ref{PFM}. In the present work, the grains and GBs maintain an isotropic diffusion behaviour. The contribution of crystallographic orientation and GB angle (high and low-angle GBs) will be addressed in future work. 

For the underlying corrosion mechanism considered, the total free energy functional of the heterogeneous system in Fig. \ref{pf_description} can be expressed as
\begin{equation}
    \mathcal{F} = \int_\Omega[f^\mathrm{chem}(\overline{c},\phi) + f^\mathrm{grad}(\nabla\phi)]d\Omega,
    \label{eq1}
\end{equation}
where $f^\mathrm{chem}$ and $f^\mathrm{grad}$ are the chemical and interfacial free energy densities detailed further below. $\Omega$ in the previous expression represents the whole system domain that includes both the corrosive liquid Li environment and the polycrystalline material. 

The chemical free energy density is expressed as the weighted sum of free energy density from each contributing phase \cite{Wheeler1992Phase-fieldAlloys}
\begin{equation}
    f^\mathrm{chem}(\overline{c},\phi) = h(\phi) f^\mathrm{chem}_\mathrm{s}(\overline{c}_\mathrm{s}) + (1-h(\phi)) f^\mathrm{chem}_\mathrm{l}(\overline{c}_{\mathrm{l}}) + \omega g(\phi),
    \label{eq2}
\end{equation}
where $f^\mathrm{chem}_\mathrm{s}(\overline{c}_\mathrm{s})$ and $f^\mathrm{chem}_\mathrm{l}(\overline{c}_\mathrm{l})$ are the chemical free energy densities with respect to the normalized concentrations in the liquid ($\overline{c}_\mathrm{l}$) and solid ($\overline{c}_\mathrm{s})$ phases. $g(\phi)=16\phi^2(1-\phi)^2$ is the double-well free energy function employed to describe the two equilibrium states for the solid ($\phi = 1$) and the liquid ($\phi = 0$) phases. $\omega$ is the constant that determines the height of the energy barrier at $\phi = 1/2$ between the two minima at $\phi = 0$ and $\phi = 1$. $h(\phi)=\phi^3(6\phi^2-15\phi+10)$ is a monotonously increasing interpolation function that interpolates the chemical free energy density between the two phases. The chemical free energy density of each phase can be reasonably approximated by a simple parabolic function \cite{Hu2007ThermodynamicApproach} around equilibrium concentrations with the same free energy density curvature parameter $A$ as \cite{Makuch2024ACracking}
\begin{equation}
    f^\mathrm{chem}_\mathrm{s}(\overline{c}_\mathrm{s}) = \frac{1}{2} A (\overline{c}_\mathrm{s}-\overline{c}_\mathrm{s,eq})^2
    \quad\quad
    f^\mathrm{chem}_\mathrm{l}(\overline{c}_\mathrm{l}) = \frac{1}{2} A (\overline{c}_\mathrm{l}-\overline{c}_\mathrm{l,eq})^2,
    \label{eq5}
\end{equation}
where $\overline{c}_\mathrm{s,eq} = c_\mathrm{solid} / c_\mathrm{solid}=1$ and $\overline{c}_\mathrm{l,eq} = c_\mathrm{sat} / c_\mathrm{solid}$  are the normalized equilibrium concentrations for the solid and liquid phase. Here, $c_\mathrm{sat}$ represents the saturation limit of the metal species in the liquid phase. Each material point in the present model is characterized as a mixture of both solid and liquid phases with different compositions yet the same diffusion chemical potentials \cite{Kim1999Phase-fieldAlloys}. This assumption renders the following expression for the chemical free energy density $f^\mathrm{chem}(\overline{c},\phi)$ \cite{Makuch2024ALayer}  
\begin{equation}
    f^\mathrm{chem}(\overline{c},\phi)=\frac{1}{2} A [\overline{c}-h(\phi)(\overline{c}_\mathrm{s,eq} - \overline{c}_\mathrm{l,eq}) - \overline{c}_\mathrm{l,eq}]^2 + \omega g(\phi).
    \label{fchem}
\end{equation}
 
The interfacial free energy density is commonly expressed as
 \begin{equation}
     f^\mathrm{grad}(\nabla\phi)=\frac{1}{2}\kappa|\nabla\phi|^2
     \label{eq7}
 \end{equation}
where $\kappa$ is the isotropic gradient energy coefficient. The phase-field parameters $\omega$ in Eq. (\ref{eq2}) and $\kappa$ in Eq. (\ref{eq7}) are connected to the interfacial energy $\Gamma$ and the chosen nominal interface thickness $\ell$ as \cite{Kovacevic2020InterfacialCeramics}
\begin{equation} \label{PFparameters}
    \kappa=\frac{3}{2}\Gamma \ell
    \quad\quad
    \omega=\frac{3\Gamma}{4 \ell}.
\end{equation}

The governing equations for the independent kinematic fields $\phi(\mathbf{x}, t)$ and $\overline{c}(\mathbf{x}, t)$ are derived by minimizing the total energy of the system and conserving the total concentration of Cr composition within the system \citep{Kovacevic2023Phase-fieldApplications}. The evolution of the non-conserved phase-field parameter $\phi$ follows the Allen-Cahn equation \cite{Allen1979ACoarsening}
\begin{equation}
    \frac{\partial\phi}{\partial t}=-L\frac{\delta \mathcal{F}}{\delta\phi}=-L\Big( \frac{\partial f^\mathrm{chem}}{\partial\phi}-\kappa\nabla^2\phi \Big) \quad \text{in}\quad\Omega \qquad \kappa \mathbf{n} \cdot \nabla \phi = 0 \quad \text{on}\quad\partial\Omega,
    \label{eqnAC}
\end{equation}
where $L$ is the kinetic coefficient that characterizes the interfacial mobility and controls the motion of the solid$-$liquid interface. The magnitude of this parameter determines the underlying corrosion mechanisms, such as activation-controlled and diffusion-controlled, and regulates the corrosion rate \cite{Cui2023Electro-chemo-mechanicalImplementation, Makuch2024ACracking, Ansari2019ModelingMetals, Kandekar2024MasteringScheme, Mai2018NewProcesses}. The condition for $L$ to capture both mechanisms is given in the following section. 

The transport of Cr composition in the system is subjected to the conservation law
\begin{equation} \label{diffusion}
    \frac{\partial\overline{c}}{\partial t} = - \nabla\cdot \mathbf{J} \quad\mathrm{in}\;\Omega \qquad  
\mathbf{J} = - M \nabla \Big(\frac{\delta \mathcal{F} }{\delta \overline{c}} \Big) \qquad \mathbf{n} \cdot \mathbf{J} = 0 \quad \text{on}\quad\partial\Omega,
\end{equation}
where $\mathbf{J}$ stands for the diffusional flux and $M$ the mobility parameter that characterizes the motion of Cr composition. Here, the mobility parameter is expressed as: $M = D/(\partial^2 f^\mathrm{chem}/\partial \overline{c}^2) = D/A$ \cite{Makuch2024ACracking}, where $D$ is the effective diffusion coefficient of Cr composition. After substituting the expression for the mobility parameter into Eq. (\ref{diffusion}), the diffusional flux is written as  
\begin{equation}
    \mathbf{J} = -D \nabla \overline{c} - D h^\prime(\phi)(\overline{c}_\mathrm{l,eq}-\overline{c}_\mathrm{s,eq}) \nabla \phi.
    \label{eqnflux}
\end{equation}
The effective diffusion coefficient of Cr composition is interpolated between the grain bulk and GBs as
\begin{equation}
    D = D_\mathrm{gb}^\prime \eta + (1-\eta) D_\mathrm{mg} \quad\quad  D^{\prime}_\mathrm{gb}=\frac{\delta_{gb}}{l_p}D_\mathrm{gb},
    \label{eqngbdiff}
\end{equation}
where $D_\mathrm{mg}$ and $D_\mathrm{gb}$ are the diffusion coefficient of Cr composition in the grain bulk and along GBs, respectively. $D_\mathrm{gb} \gg D_\mathrm{mg}$ is enforced to enhance the diffusion of Cr composition along GBs. $D^{\prime}_\mathrm{gb}$ in the previous equation is the effective diffusion coefficient of Cr composition along GBs defined using the constant product approach \cite{Simon2022MechanisticParticles}. This approach relates the actual diffusion coefficient $D_\mathrm{gb}$ and the ratio between the physical $\delta_{gb}$ and computational thickness $l_p$ of the Cr depletion region along GBs. The expression proportionally alters the GB diffusivity with respect to an experimentally determined GB thickness, which in turn, rectifies the relative contribution of the otherwise unaltered GB diffusivity and additionally decreases the computational expense of simulating nanometre-sized features. The governing equation for the interpolating parameter $\eta$ is given as \cite{Nguyen2016AMicrotomography}
\begin{equation}
    \nabla\cdot(-l_p^2\nabla\eta)+\eta=0 \quad \text{in}\quad\Omega \qquad l_p^2 \mathbf{n} \cdot \nabla \eta = 0 \quad \text{on}\quad\partial\Omega \quad \text{and} \quad \eta = 1 \quad \text{on GBs}.
    \label{eqngbinter}
\end{equation}
GBs are given a value of $\eta=1$ and smoothly transitions to $\eta\rightarrow 0$ further away from the GBs, Fig.  \ref{pf_description}. The sensitivity of the evolution of IGC to the computational thickness of the smeared GB region is investigated in Section \ref{lp_results}. It is assumed in the present formulation that the Cr GB depletion region is given a uniform thickness, neglecting any confinement effects.

The framework developed is implemented into the multi-purpose finite element software package COMSOL Multiphysics \cite{COMSOLSweden}. The computational domain is discretized using triangular finite elements with second-order Lagrangian interpolation functions. All regions expected to corrode in the metal grains are given a characteristic maximum element size at least five times smaller than the interfacial thickness $\ell$ to ensure a smooth transition between the metal and corrosive agent. This setting proved sufficient based on previous literature \cite{Makuch2024ACracking, Makuch2024ALayer, Kovacevic2023Phase-fieldApplications}. Moreover, as the evolution of the interface is expected to be most prominent at the GBs compared to the metal grain, a maximum element size of $\ell$/20 is applied to all GBs. Finally, to limit the computational cost, the remaining domain of the solid phase is given a maximum element size of $\ell$. These conditions are fulfilled in all the simulations. The depictions of the finite element mesh for a representative case study are shown in Fig. \ref{comp_domain_exp_setup}(c) and Fig. \ref{comp_domain_exp_setup}(d). Each simulation consists of a two-step study. The governing equation (\ref{eqngbinter}) for the interpolating parameter $\eta$ that defines the smeared GB thickness is solved in the first step using a steady-state (time-independent) solver. The governing equations (\ref{eqnAC}) and (\ref{diffusion}) for the evolution of the phase-field parameter $\phi$ and Cr composition $\overline{c}$ are then solved in the second step using a time-dependent study. An implicit time-stepping method is used for temporal discretization in the time-dependent step. A fully coupled solution algorithm is selected to solve the governing equations. The maximum time step is 2 hours. The solver accuracy in each time step is controlled by a relative tolerance of 10$^{-4}$. The code developed together with example case studies and documentation are available at \url{https://mechmat.web.ox.ac.uk/codes} (after article acceptance).

\subsection{Model calibration and validation} \label{setup}

\begin{figure}[h!]
    \centering
    \includegraphics[width = 14 cm]{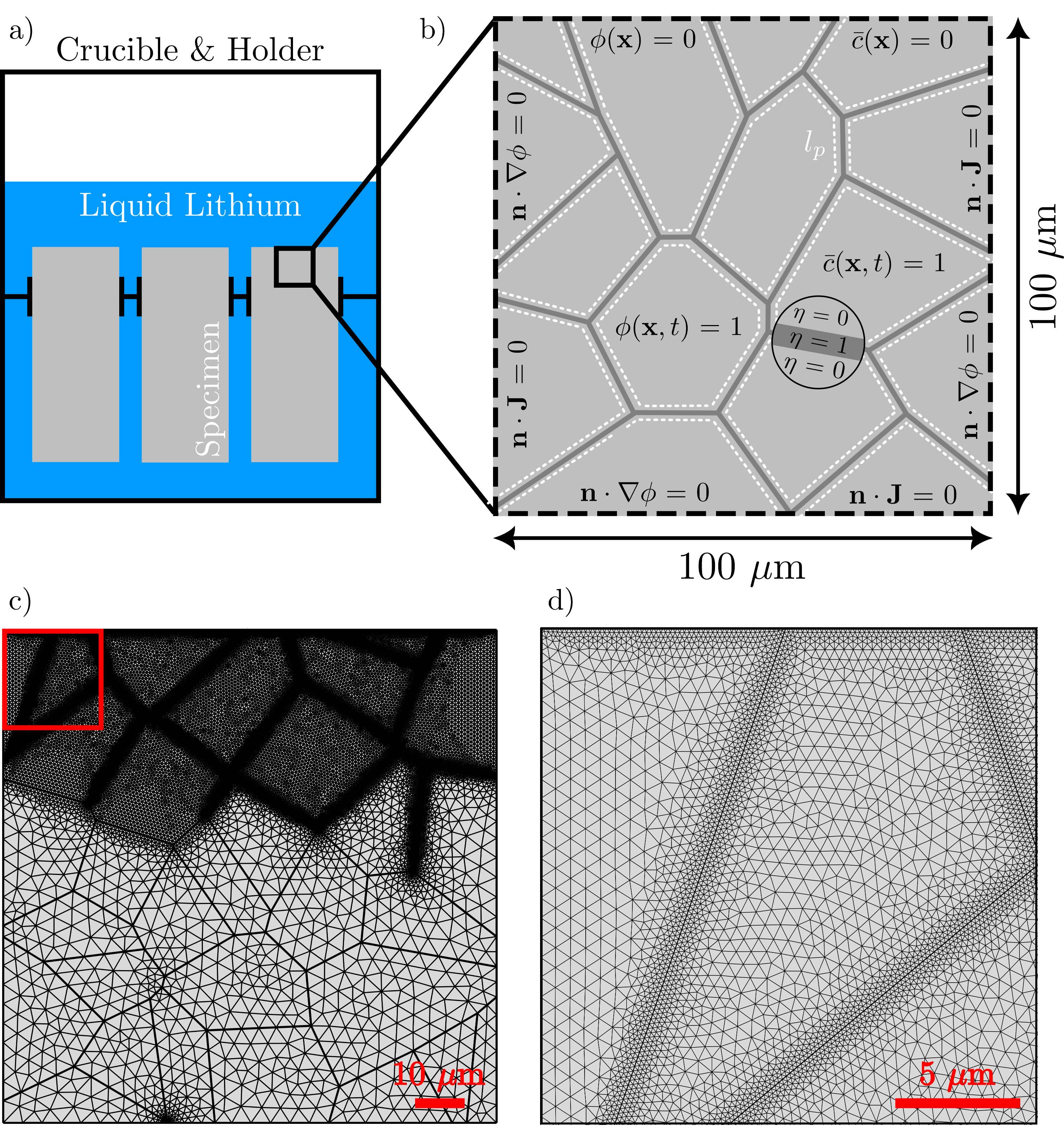}
    \captionsetup{labelfont = bf,justification = raggedright}
    \caption{\textbf{Computational conditions, domain and meshing.} Schematic of the \textbf{a} experimental apparatus used by Xu et al. \cite{Xu2008CorrosionConditions} to expose F/M specimen to static liquid Li, \textbf{b} corresponding computational domain consisting of a polycrystalline material with the initial values and boundary conditions, \textbf{c} finite element mesh of the whole computational domain, and \textbf{d} enlarged area corresponding to the red square in \textbf{c} highlighting the finite element size within the grains and along the GBs in the expected area of interface propagation. The exposed surface is the upper most boundary.}
    \label{comp_domain_exp_setup}
\end{figure}

The model developed is calibrated and validated against experimental data given in \cite{Xu2008CorrosionConditions} where greater detail regarding the experimental set-up can be found. The steel specimen underwent a normalized and tempered heat treatment resulting in a tempered martensite microstructure with a prior austenitic grain (PAG) size of 20 $\upmu$m \cite{Tsisar2011StructuralCoarsening}. The experimental apparatus used by Xu \textit{et al.} is displayed in Fig. \ref{comp_domain_exp_setup}. 100 mL of high purity Li was used, whereby the concentration of nitrogen was estimated to never surpass 100 ppm. Nitrogen has been experimentally reported to exacerbate the corrosion process through chemical reactions producing the stable ternary nitride corrosion product, thereby intensifying the leaching and degradation of structural materials \cite{Tsisar2011EffectLithium, Coen1984CompatibilityIspra}. Nonetheless, following the equilibrium concentration of nitrogen required to form the ternary nitride corrosion complex \cite{Knaster2017AssessmentStudy}, the propensity for the corrosion complex to have nucleated and thus contribute to the penetration process was unlikely, consolidating the bulk diffusion mechanism adopted herein. The geometric ratio between the volume of Li to the total surface area of the specimen was $\sim$ 4 cm. The specimen was exposed to static liquid Li at 600 $\degree$C for 750 hours. Experimental measurements in terms of weight loss and corrosion depth are used to calibrate the model.

The numerical simulation is conducted by considering a 2D computational domain with 100 $\upmu$m in length and height. Only 2D simulations are performed for simplicity whereby the implications of this assumption are discussed later. The dimensions are selected based on minimizing computational cost while maintaining an adequate region of microstructural features to observe its influence. The computational domain is characterized by a microstructure with an average grain size equal to the PAG size, 20 $\upmu$m from the experiment in \cite{Xu2008CorrosionConditions}, Fig. \ref{comp_domain_exp_setup}. Solely simulating prior austenitic GBs is a reasonable approximation based on literature that details the extensive corrosion and significant quantity of liquid Li at high-angle GBs following compatibility experiments \cite{Coen1984CompatibilityIspra, Ruedl1983EffectSteel}. Nonetheless, incorporating the entire tempered martensitic constituents (i.e., lath and block boundaries) is important to fully capture the microstructural effects of liquid Li IGC and, as such, this will be investigated in the future. Ten different microstructures are constructed to have a statistically significant sample and generate an average. The ten microstructures used can be viewed in Fig. S.1 (Supplementary Information). The simulation starts by solving for the smeared GB thickness using the governing equation and corresponding boundary conditions in Eq. (\ref{eqngbinter}). $\eta=1$ is enforced on selected GBs that are expected to corrode. Afterward, the resulting set of the governing equations (\ref{eqnAC}) and (\ref{diffusion}) is solved with accompanying initial and boundary conditions, Fig. \ref{comp_domain_exp_setup}. The computational domain initially consists solely of the metal phase. Thus, the initial values for the phase-field variable and Cr concentration are set to $\phi = 1$ and $\bar{c} = 1$. The liquid Li phase is represented through a concentration sink implemented by prescribing the Cr concentration to zero. The sink is positioned on the uppermost boundary of the simulated domain. As the simulated solid phase is only exposed from the upper boundary, the other three edges possess zero flux boundary conditions ($\mathbf{n} \cdot \nabla \phi = 0$ and $\mathbf{n} \cdot \mathbf{J} = 0$) for both phase-field variable and Cr concentration. Substituting the liquid phase with a concentration sink does discount the effect of saturation, which ultimately halts the corrosion process in static conditions \cite{Xu2008CorrosionConditions}. Consequentially, using a concentration sink to approximate the liquid phase facilitates endless IGC. Yet, it is a justifiable approximation before saturation is reached.

The material properties and model parameters employed in the simulation are summarized in Table \ref{parameters}. 
The diffusion coefficients of Cr in the metal grain bulk $D_\mathrm{mg}$ and along GBs $D_\mathrm{gb}$ are taken from Čermák \textit{et al.} \cite{Cermak1996Low-temperatureAlloys}. Due to a lack of experimental work, the interfacial energy $\Gamma$ is obtained based on GB interfacial energies for BCC metals and alloys \cite{Li2023TheoreticalMetals, Scheiber2016AbMetals}. A value of $\Gamma$ = 4 J/m$^2$ is utilized in the present work. The interfacial thickness $\ell$ is selected to be significantly smaller than the computational domain and characteristic grain size ($\ell$ = 4 $\upmu$m). The phase-field parameters $\omega$ (Eq. (\ref{fchem})) and $\kappa$ (Eq. (\ref{eq7})) are then expressed using $\ell$ and $\Gamma$, as indicated in Eq. (\ref{PFparameters}). The thickness of the Cr-depleted region $\delta_{gb}$ is adopted based on experimental work \cite{Nakamichi2008QuantitativeFE-STEM}, which found that the thickness varied between 10$-$15 nm. As such, a conservative approach is taken in the present study, setting $\delta_{gb}$ = 15 nm. The computational GB thickness $l_p$ is set to $l_p$ = 100 nm. The two aforementioned thicknesses; $\delta_{gb}$ represents the experimentally determined parameter and $l_p$ signifies the thickness of the simulated GB region, are employed via the constant product approach, to link the artificial microstructure to physical material properties. The chemical free energy density curvature parameter $A$ in Eq. (\ref{fchem}) is chosen from similar studies \cite{Makuch2024ACracking, Makuch2024ALayer, Kovacevic2023Phase-fieldApplications, Cui2023Electro-chemo-mechanicalImplementation}. Employing the first-principles calculations and the CALPHAD approach can yield a more accurate description of the free energy functional with thermodynamic properties over various compositions and temperatures \cite{KumarThakur2023ASystems}, which can be fed into the current phase-field framework. The chemical driving force in the present model originates from the difference in Cr composition and the equilibrium values $c_{\mathrm{solid}}$ and $c_{\mathrm{sat}}$ in each phase, Eq. (\ref{fchem}). The former represents the initial known concentration of Cr in a steel specimen before exposure to liquid Li, $c_\mathrm{solid}$ = 13.4 mol/L \cite{Xu2008CorrosionConditions}. The latter signifies the equilibrium concentration of Cr in the steel specimen following exposure to liquid Li under static conditions. Due to the extended exposure time tested in the compatibility experiments conducted by Xu \textit{et al.} \cite{Xu2008CorrosionConditions}, along with the knowledge that chemical equilibrium in static conditions with 100 mL of liquid Li is reached within 250 h \cite{Xu2007CompatibilityLithium, Xu2008CorrosionConditions}, it is warranted to take the post-corrosion test surface concentration of Cr as the equilibrium value, $c_{\mathrm{sat}}$ = 10.3 mol/L \cite{Tsisar2011StructuralCoarsening}. The interface kinetics coefficient $L$ governs the motion of the metal$-$liquid interface and, as such, dictates the rate-limiting step of the corrosion process. Comparatively large values give rise to diffusion-controlled corrosion and transitions to activation-controlled as the value decreases \cite{Cui2023Electro-chemo-mechanicalImplementation}. Dimensional analysis \cite{Kovacevic2023Phase-fieldApplications} indicates that $L$ needs to be $L \gg D_\mathrm{gb}/(\ell^2 \omega)$ for diffusion-controlled and $L \ll D_\mathrm{gb}/(\ell^2 \omega)$ for activation-controlled regimes. To ensure the corrosion process possesses adequate kinetics to operate under diffusion control, it is given a value of $L = $ 1 m$^2$/(N$\cdot$s), which satisfies the previous condition.

\begin{figure}[h!]
    \centering
    \includegraphics[width=16cm]{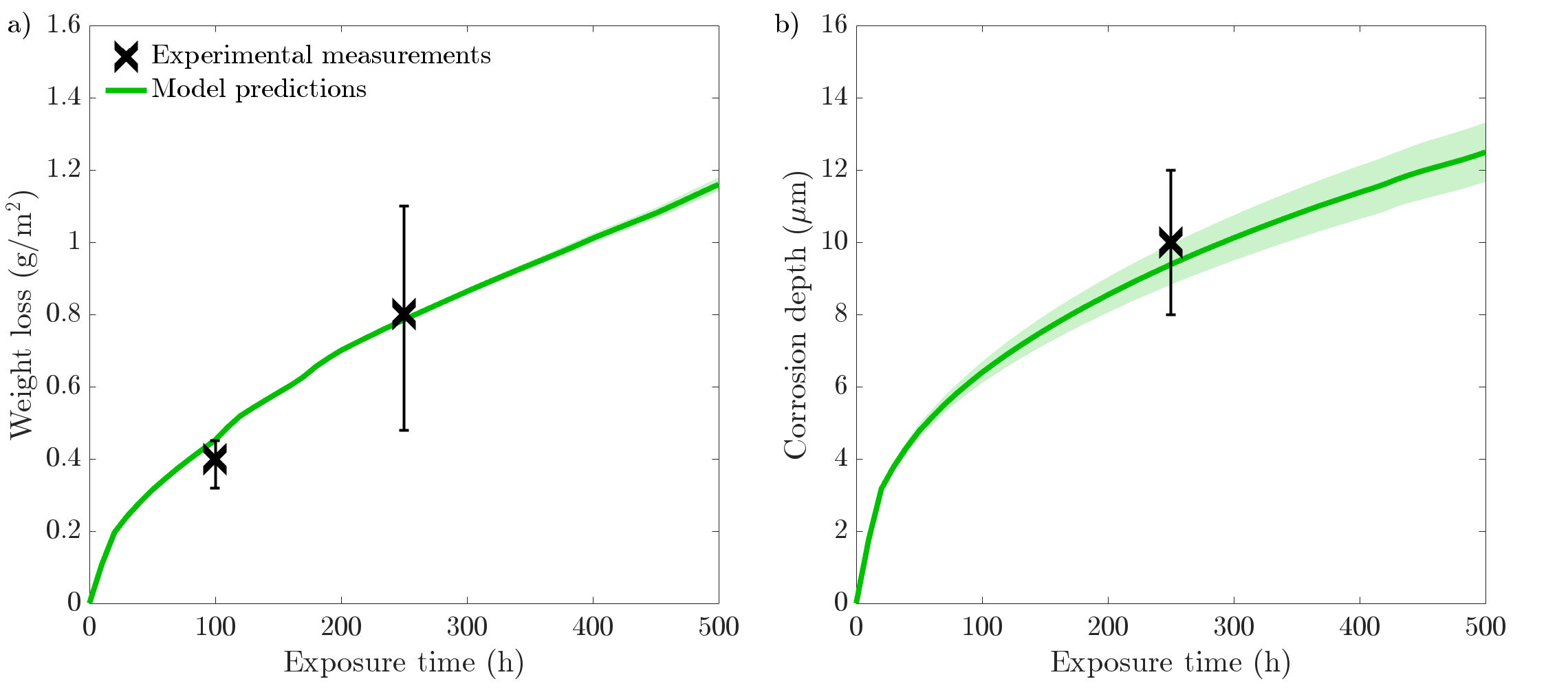}
    \captionsetup{labelfont = bf,justification = raggedright}
    \caption{\textbf{Phase-field predictions following a 500-hour simulation with a 20 $\upmu$m average grain size microstructure compared to experimental measurements from Xu \textit{et al.} \cite{Xu2008CorrosionConditions}.} Phase-field predictions for \textbf{a} weight loss and \textbf{b} corrosion depth. The data presented is taken from an average of ten microstructures where the shaded coloured region represents the SD of the ten simulations.}
    \label{simvexp}
\end{figure}

The weight loss of the simulated specimen is determined in the simulation as
\begin{equation}
    \Delta m=\int_{0}^{\overline{c}_{\mathrm{l,eq}}} \overline{c} \cdot c_\mathrm{solid} \cdot A_{Cr} \, d\Omega,
    \label{weightlosseqn}
\end{equation}
where $\Delta m$ represents the mass loss in grams and $A_{Cr}$ is the atomic weight of Cr in grams/mole. In accordance with the experimental procedure in evaluating the susceptibility of specimens with liquid Li, the obtained weight loss is normalized with respect to the exposed surface area of the specimen. As such, Eq. (\ref{weightlosseqn}) is divided by the length of the concentration sink region, 100 $\upmu$m. The corrosion depth is determined by taking the mean corrosion depth from all advancing corrosion fronts ($\phi<0.5$) along GBs that are diffusing deeper towards the bulk. Fig. \ref{simvexp} illustrates the weight loss and corrosion depth averaged from ten varying microstructures. Fig. \ref{simvexp}(a) showcases the strong resemblance between the phase-field predictions and experimental data regarding weight loss. The weight loss predicted by the model passes through the upper relative error at 100 hours and matches the experimental data at 250 hours with high accuracy. Moreover, the corrosion predictions displayed in Fig. \ref{simvexp}(b) similarly shows a high correlation to the experimental data, resulting in a corrosion depth of 9.4 $\upmu$m after 250 hours. After 500 hours, the model predicted an average weight loss of 1.16 g/m$^2$ where the liquid Li reached an average depth of 12.5 $\upmu$m. The logarithmically varying profile for the weight loss and corrosion depth signifies that the corrosion process is in the diffusion-controlled region, further aligning with the literature and proving that the phase-field mobility parameter $L$ is adequately selected. The close comparison between the phase-field predictions and experimental measurements \cite{Xu2008CorrosionConditions} showcases the capabilities of the current model to simulate IGC. Following model calibration and validation, microstructural and GB properties are altered to observe their dependence on the corrosion resistance of the simulated F/M specimen to static liquid Li in Section \ref{results}. The data is compared against the reference geometry described in this section. 

\section*{Data Availability} \label{data_ava}
The relevant data are available from the corresponding authors upon reasonable request.

\section*{Code Availability} \label{code_ava}
The code developed together with example case studies and documentation is available at \url{https://mechmat.web.ox.ac.uk/codes}.

\section*{Acknowledgements}
This work is supported by the EPSRC Centre for Doctoral Training in Nuclear Energy Futures and the agreement between UKAEA and Imperial College London (EP/S023844/1). D.N.-M. work has been carried out within the framework of the EUROfusion Consortium, funded by the European Union via the Euratom Research and Training Programme (101052200 — EUROfusion). The work at UKAEA was partially supported by the Broader Approach Phase II agreement under the PA of IFERC2-T2PA02. Views and opinions expressed are however those of the author(s) only and do not necessarily reflect those of the European Union or the European Commission. Neither the European Union nor the European Commission can be held responsible for them. D.N.-M. and J.L. also acknowledge funding by the EPSRC Energy Programme (EP/W006839/1). S.K. and E.M.-P. acknowledge financial support from UKRI’s Future Leaders Fellowship program [Grant MR/V024124/1].

\section*{Author Contributions}
\textbf{A.L.:} Writing – original draft, Visualization, Validation, Software, Methodology, Investigation, Formal analysis, Data curation, Conceptualization. \textbf{S.K.:} Writing – review \& editing,
Supervision, Software, Methodology, Investigation, Formal analysis,
Conceptualization. \textbf{D.N.-M.:} Writing – review \& editing, Supervision, Project administration, Funding acquisition, Conceptualization. \textbf{J.L.:} Writing – review \& editing, Supervision, Project administration, Funding acquisition, Conceptualization. \textbf{E.M.-P.:} Writing – review \& editing, Supervision, Software, Resources, Project administration, Methodology, Funding acquisition, Conceptualization. \textbf{M.W.:} Writing – review \& editing, Supervision, Project administration, Funding acquisition, Conceptualization. All authors have read and approved the manuscript.

\section*{Competing Interests}
The authors declare no competing interests.

\begin{singlespace}

\end{singlespace}

\begin{table}
    \captionsetup{labelfont = bf,justification = raggedright}
    \caption{\textbf{Material and computational parameters employed in the model.}}
    \centering
    \begin{tabular}{lccc}
    \toprule
    Parameter & Value & Unit & Ref \\
    \midrule
    Chemical free energy density curvature parameter $A$ & $5\times10^9$ & N/m$^2$\\
    Interface kinetics coefficient $L$ & 1 & m$^2$/(N$\cdot$s)\\
    Concentration in the solid phase $c_\mathrm{solid}$ & 13.4 & mol/L & \cite{Xu2008CorrosionConditions} \\
    Saturated concentration in the liquid phase $c_\mathrm{sat}$ & 10.3 & mol/L & \cite{Tsisar2011StructuralCoarsening} \\
    Interfacial energy $\Gamma$ & 4 & N/m & \cite{Scheiber2016AbMetals, Li2023TheoreticalMetals} \\
    Computational GB thickness $l_p$ & 100 & nm \\
    Physical Cr depletion thickness $\delta_{gb}$ & 15 & nm & \cite{Nakamichi2008QuantitativeFE-STEM} \\
    Interfacial thickness $\ell$ & 4 & $\upmu$m \\
    Grain boundary diffusivity $D_\mathrm{gb}$ & $1.70\times 10^{-15}$ & m$^2$/s & \cite{Cermak1996Low-temperatureAlloys} \\
    Metal grain diffusivity $D_\mathrm{mg}$ & $5.11\times10^{-21}$ & m$^2$/s & \cite{Cermak1996Low-temperatureAlloys} \\
    \bottomrule
    \end{tabular}
    \label{parameters}
\end{table}


\small
\begin{thebibliography}{10}
\expandafter\ifx\csname url\endcsname\relax
  \def\url#1{\burl{#1}}\fi
\expandafter\ifx\csname urlprefix\endcsname\relax\def\urlprefix{URL }\fi
\providecommand{\bibinfo}[2]{#2}
\providecommand{\eprint}[2][]{\url{#2}}
\providecommand{\doi}[1]{\url{https://doi.org/#1}}


\bibitem{Smith1981BlanketReactors}
\bibinfo{author}{Smith, D.}
\newblock \bibinfo{title}{{Blanket materials for dt fusion reactors}}.
\newblock \emph{\bibinfo{journal}{Journal of Nuclear Materials}}
  \textbf{\bibinfo{volume}{103}}, \bibinfo{pages}{19--29}
  (\bibinfo{year}{1981}).

\bibitem{Shanliang2003NeutronicMaterials}
\bibinfo{author}{Shanliang, Z.} \& \bibinfo{author}{Yican, W.}
\newblock \bibinfo{title}{{Neutronic Comparison of Tritium-Breeding Performance
  of Candidate Tritium-Breeding Materials}}.
\newblock \emph{\bibinfo{journal}{Plasma Science and Technology}}
  \textbf{\bibinfo{volume}{5}}, \bibinfo{pages}{1995--2000}
  (\bibinfo{year}{2003}).

\bibitem{Malang1995ComparisonBlankets}
\bibinfo{author}{Malang, S.} \& \bibinfo{author}{Mattas, R.}
\newblock \bibinfo{title}{{Comparison of lithium and the eutectic lead-lithium
  alloy, two candidate liquid metal breeder materials for self-cooled
  blankets}}.
\newblock \emph{\bibinfo{journal}{Fusion Engineering and Design}}
  \textbf{\bibinfo{volume}{27}}, \bibinfo{pages}{399--406}
  (\bibinfo{year}{1995}).

\bibitem{Kirillov2006RFTests}
\bibinfo{author}{Kirillov, I.}, \bibinfo{author}{Shatalov, G.} \&
  \bibinfo{author}{Strebkov, Y.}
\newblock \bibinfo{title}{{RF TBMs for ITER tests}}.
\newblock \emph{\bibinfo{journal}{Fusion Engineering and Design}}
  \textbf{\bibinfo{volume}{81}}, \bibinfo{pages}{425--432}
  (\bibinfo{year}{2006}).

\bibitem{Bloom1998TheSystems}
\bibinfo{author}{Bloom, E.~E.}
\newblock \bibinfo{title}{{The challenge of developing structural materials for
  fusion power systems}}.
\newblock \emph{\bibinfo{journal}{Journal of Nuclear Materials}}
  \textbf{\bibinfo{volume}{258-263}}, \bibinfo{pages}{7--17}
  (\bibinfo{year}{1998}).

\bibitem{Giancarli2006BreedingDEMO}
\bibinfo{author}{Giancarli, L.} \emph{et~al.}
\newblock \bibinfo{title}{{Breeding Blanket Modules testing in ITER: An
  international program on the way to DEMO}}.
\newblock \emph{\bibinfo{journal}{Fusion Engineering and Design}}
  \textbf{\bibinfo{volume}{81}}, \bibinfo{pages}{393--405}
  (\bibinfo{year}{2006}).

\bibitem{Smith1998MaterialsSystems}
\bibinfo{author}{Smith, D.}, \bibinfo{author}{Billone, M.},
  \bibinfo{author}{Majumdar, S.}, \bibinfo{author}{Mattas, R.} \&
  \bibinfo{author}{Sze, D.-K.}
\newblock \bibinfo{title}{{Materials integration issues for high performance
  fusion power systems}}.
\newblock \emph{\bibinfo{journal}{Journal of Nuclear Materials}}
  \textbf{\bibinfo{volume}{258-263}}, \bibinfo{pages}{65--73}
  (\bibinfo{year}{1998}).

\bibitem{Giancarli2020OverviewActivities}
\bibinfo{author}{Giancarli, L.~M.} \emph{et~al.}
\newblock \bibinfo{title}{{Overview of recent ITER TBM Program activities}}.
\newblock \emph{\bibinfo{journal}{Fusion Engineering and Design}}
  \textbf{\bibinfo{volume}{158}}, \bibinfo{pages}{111674}
  (\bibinfo{year}{2020}).

\bibitem{Klueh2005ElevatedReactors}
\bibinfo{author}{Klueh, R.~L.}
\newblock \bibinfo{title}{{Elevated temperature ferritic and martensitic steels
  and their application to future nuclear reactors}}.
\newblock \emph{\bibinfo{journal}{International Materials Reviews}}
  \textbf{\bibinfo{volume}{50}}, \bibinfo{pages}{287--310}
  (\bibinfo{year}{2005}).

\bibitem{Chopra1984CompatibilityReactors}
\bibinfo{author}{Chopra, O.~K.} \& \bibinfo{author}{Tortorelli, P.~F.}
\newblock \bibinfo{title}{{Compatibility of materials for use in liquid-metal
  blankets of fusion reactors}}.
\newblock \emph{\bibinfo{journal}{Journal of Nuclear Materials}}
  \textbf{\bibinfo{volume}{123}}, \bibinfo{pages}{1201--1212}
  (\bibinfo{year}{1984}).

\bibitem{Tortorelli1984LiquidDevelopment}
\bibinfo{author}{Tortorelli, P.~F.} \& \bibinfo{author}{DeVan, J.~H.}
\newblock \bibinfo{editor}{{Oak Ridge National Lab}} (ed.)
  \emph{\bibinfo{title}{{Liquid metal corrosion considerations in alloy
  development}}}.
\newblock (ed.\bibinfo{editor}{{Oak Ridge National Lab}})
  (\bibinfo{address}{Tennessee, United States}, \bibinfo{year}{1984}).

\bibitem{Tortorelli1988CorrosionReactions}
\bibinfo{author}{Tortorelli, P.~F.}
\newblock \bibinfo{title}{{Corrosion of ferritic steels by molten lithium:
  Influence of competing thermal gradient mass transfer and surface product
  reactions}}.
\newblock \emph{\bibinfo{journal}{Journal of Nuclear Materials}}
  \textbf{\bibinfo{volume}{155-157}}, \bibinfo{pages}{722--727}
  (\bibinfo{year}{1988}).

\bibitem{Lyublinski1995NumericalEutectic}
\bibinfo{author}{Lyublinski, I.}, \bibinfo{author}{Evtikhin, V.},
  \bibinfo{author}{Pankratov, V.} \& \bibinfo{author}{Krasin, V.}
\newblock \bibinfo{title}{{Numerical and experimental determination of metallic
  solubilities in liquid lithium, lithium-containing nonmetallic impurities,
  lead and lead-lithium eutectic}}.
\newblock \emph{\bibinfo{journal}{Journal of Nuclear Materials}}
  \textbf{\bibinfo{volume}{224}}, \bibinfo{pages}{288--292}
  (\bibinfo{year}{1995}).

\bibitem{Chopra1988CompatibilitySystems}
\bibinfo{author}{Chopra, O.~K.} \& \bibinfo{author}{Smith, D.~L.}
\newblock \bibinfo{title}{{Compatibility of ferritic steels in forced
  circulation lithium and Pb-17Li systems}}.
\newblock \emph{\bibinfo{journal}{Journal of Nuclear Materials}}
  \textbf{\bibinfo{volume}{155-157}}, \bibinfo{pages}{715--721}
  (\bibinfo{year}{1988}).

\bibitem{Tortorelli1981CorrosionApplications}
\bibinfo{author}{Tortorelli, P.} \& \bibinfo{author}{Chopra, O.}
\newblock \bibinfo{title}{{Corrosion and compatibility considerations of liquid
  metals for fusion reactor applications}}.
\newblock \emph{\bibinfo{journal}{Journal of Nuclear Materials}}
  \textbf{\bibinfo{volume}{103}}, \bibinfo{pages}{621--632}
  (\bibinfo{year}{1981}).

\bibitem{Ruedl1982IntergranularSteels}
\bibinfo{author}{Ruedl, E.}, \bibinfo{author}{Coen, V.},
  \bibinfo{author}{Sasaki, T.} \& \bibinfo{author}{Kolbe, H.}
\newblock \bibinfo{title}{{Intergranular lithium penetration of low-Ni, Cr-Mn
  Austenitic stainless steels}}.
\newblock \emph{\bibinfo{journal}{Journal of Nuclear Materials}}
  \textbf{\bibinfo{volume}{110}}, \bibinfo{pages}{28--36}
  (\bibinfo{year}{1982}).

\bibitem{Bell1989TheLithium}
\bibinfo{author}{Bell, G.~E.} \& \bibinfo{author}{Abdou, M.~A.}
\newblock \bibinfo{title}{{The Role of Carbides in the Corrosion Behavior of
  Fe-12Cr-1MoVW Steel in Liquid Lithium}}.
\newblock \emph{\bibinfo{journal}{Fusion Technology}}
  \textbf{\bibinfo{volume}{15}}, \bibinfo{pages}{315--320}
  (\bibinfo{year}{1989}).

\bibitem{Hosaka2022ChemicalNaK}
\bibinfo{author}{Hosaka, T.}, \bibinfo{author}{Kondo, M.},
  \bibinfo{author}{Sato, S.}, \bibinfo{author}{Ando, M.} \&
  \bibinfo{author}{Nozawa, T.}
\newblock \bibinfo{title}{{Chemical compatibility of F82H and 316L in liquid
  metal heat transfer mediums Li, Na and NaK}}.
\newblock \emph{\bibinfo{journal}{Journal of Nuclear Materials}}
  \textbf{\bibinfo{volume}{561}}, \bibinfo{pages}{153546}
  (\bibinfo{year}{2022}).

\bibitem{Xu2007CompatibilityLithium}
\bibinfo{author}{Xu, Q.}, \bibinfo{author}{Nagasaka, T.} \&
  \bibinfo{author}{Muroga, T.}
\newblock \bibinfo{title}{{Compatibility of Low Activation Ferritic Steels with
  Liquid Lithium}}.
\newblock \emph{\bibinfo{journal}{Fusion Science and Technology}}
  \textbf{\bibinfo{volume}{52}}, \bibinfo{pages}{609--612}
  (\bibinfo{year}{2007}).

\bibitem{Tsisar2011StructuralCoarsening}
\bibinfo{author}{Tsisar, V.}, \bibinfo{author}{Kondo, M.},
  \bibinfo{author}{Muroga, T.}, \bibinfo{author}{Nagasaka, T.} \&
  \bibinfo{author}{Yeliseyeva, O.}
\newblock \bibinfo{title}{{Structural and compositional transformations in the
  near-surface layers of Fe–Cr based steels exposed to lithium – Effect of
  alloying and corrosion-assisted substructure coarsening}}.
\newblock \emph{\bibinfo{journal}{Corrosion Science}}
  \textbf{\bibinfo{volume}{53}}, \bibinfo{pages}{441--447}
  (\bibinfo{year}{2011}).

\bibitem{Zhang2024StudyLithium}
\bibinfo{author}{Zhang, D.} \emph{et~al.}
\newblock \bibinfo{title}{{Study on corrosion behavior of China low activation
  ferritic/martensitic steel in static liquid lithium}}.
\newblock \emph{\bibinfo{journal}{Nuclear Materials and Energy}}
  \textbf{\bibinfo{volume}{38}}, \bibinfo{pages}{101594}
  (\bibinfo{year}{2024}).

\bibitem{Sarkar2012ADissolution}
\bibinfo{author}{Sarkar, S.}, \bibinfo{author}{Warner, J.~E.} \&
  \bibinfo{author}{Aquino, W.}
\newblock \bibinfo{title}{{A numerical framework for the modeling of corrosive
  dissolution}}.
\newblock \emph{\bibinfo{journal}{Corrosion Science}}
  \textbf{\bibinfo{volume}{65}}, \bibinfo{pages}{502--511}
  (\bibinfo{year}{2012}).

\bibitem{Sun2014AnCorrosion}
\bibinfo{author}{Sun, W.}, \bibinfo{author}{Wang, L.}, \bibinfo{author}{Wu, T.}
  \& \bibinfo{author}{Liu, G.}
\newblock \bibinfo{title}{{An arbitrary Lagrangian–Eulerian model for
  modelling the time-dependent evolution of crevice corrosion}}.
\newblock \emph{\bibinfo{journal}{Corrosion Science}}
  \textbf{\bibinfo{volume}{78}}, \bibinfo{pages}{233--243}
  (\bibinfo{year}{2014}).

\bibitem{Jafarzadeh2019PittingModels}
\bibinfo{author}{Jafarzadeh, S.}, \bibinfo{author}{Chen, Z.},
  \bibinfo{author}{Zhao, J.} \& \bibinfo{author}{Bobaru, F.}
\newblock \bibinfo{title}{{Pitting, lacy covers, and pit merger in stainless
  steel: 3D peridynamic models}}.
\newblock \emph{\bibinfo{journal}{Corrosion Science}}
  \textbf{\bibinfo{volume}{150}}, \bibinfo{pages}{17--31}
  (\bibinfo{year}{2019}).

\bibitem{Chen2015PeridynamicDamage}
\bibinfo{author}{Chen, Z.} \& \bibinfo{author}{Bobaru, F.}
\newblock \bibinfo{title}{{Peridynamic modeling of pitting corrosion damage}}.
\newblock \emph{\bibinfo{journal}{Journal of the Mechanics and Physics of
  Solids}} \textbf{\bibinfo{volume}{78}}, \bibinfo{pages}{352--381}
  (\bibinfo{year}{2015}).

\bibitem{Gong2022NucleationAnalysis}
\bibinfo{author}{Gong, K.}, \bibinfo{author}{Wu, M.}, \bibinfo{author}{Liu, X.}
  \& \bibinfo{author}{Liu, G.}
\newblock \bibinfo{title}{{Nucleation and propagation of stress corrosion
  cracks: Modeling by cellular automata and finite element analysis}}.
\newblock \emph{\bibinfo{journal}{Materials Today Communications}}
  \textbf{\bibinfo{volume}{33}}, \bibinfo{pages}{104886}
  (\bibinfo{year}{2022}).

\bibitem{Fatoba2018SimulationApproach}
\bibinfo{author}{Fatoba, O.}, \bibinfo{author}{Leiva-Garcia, R.},
  \bibinfo{author}{Lishchuk, S.}, \bibinfo{author}{Larrosa, N.} \&
  \bibinfo{author}{Akid, R.}
\newblock \bibinfo{title}{{Simulation of stress-assisted localised corrosion
  using a cellular automaton finite element approach}}.
\newblock \emph{\bibinfo{journal}{Corrosion Science}}
  \textbf{\bibinfo{volume}{137}}, \bibinfo{pages}{83--97}
  (\bibinfo{year}{2018}).

\bibitem{Martinez-Paneda2024Phase-fieldResearch}
\bibinfo{author}{Mart{\'{i}}nez-Pa{\~{n}}eda, E.}
\newblock \bibinfo{title}{{Phase-field simulations opening new horizons in
  corrosion research}}.
\newblock \emph{\bibinfo{journal}{MRS Bulletin}} \textbf{\bibinfo{volume}{49}},
  \bibinfo{pages}{603--612} (\bibinfo{year}{2024}).

\bibitem{Cui2023Electro-chemo-mechanicalImplementation}
\bibinfo{author}{Cui, C.}, \bibinfo{author}{Ma, R.} \&
  \bibinfo{author}{Mart{\'{i}}nez-Pa{\~{n}}eda, E.}
\newblock \bibinfo{title}{{Electro-chemo-mechanical phase field modeling of
  localized corrosion: theory and COMSOL implementation}}.
\newblock \emph{\bibinfo{journal}{Engineering with Computers}}
  \textbf{\bibinfo{volume}{39}}, \bibinfo{pages}{3877--3894}
  (\bibinfo{year}{2023}).

\bibitem{Makuch2024ACracking}
\bibinfo{author}{Makuch, M.}, \bibinfo{author}{Kovacevic, S.},
  \bibinfo{author}{Wenman, M.~R.} \&
  \bibinfo{author}{Mart{\'{i}}nez-Pa{\~{n}}eda, E.}
\newblock \bibinfo{title}{{A microstructure-sensitive electro-chemo-mechanical
  phase-field model of pitting and stress corrosion cracking}}.
\newblock \emph{\bibinfo{journal}{Corrosion Science}}
  \textbf{\bibinfo{volume}{232}}, \bibinfo{pages}{112031}
  (\bibinfo{year}{2024}).

\bibitem{Ansari2019ModelingMetals}
\bibinfo{author}{Ansari, T.~Q.}, \bibinfo{author}{Luo, J.-L.} \&
  \bibinfo{author}{Shi, S.-Q.}
\newblock \bibinfo{title}{{Modeling the effect of insoluble corrosion products
  on pitting corrosion kinetics of metals}}.
\newblock \emph{\bibinfo{journal}{npj Materials Degradation}}
  \textbf{\bibinfo{volume}{3}}, \bibinfo{pages}{28} (\bibinfo{year}{2019}).

\bibitem{Lin2021Phase-fieldCracking}
\bibinfo{author}{Lin, C.} \& \bibinfo{author}{Ruan, H.}
\newblock \bibinfo{title}{{Phase-field modeling of mechano–chemical-coupled
  stress-corrosion cracking}}.
\newblock \emph{\bibinfo{journal}{Electrochimica Acta}}
  \textbf{\bibinfo{volume}{395}}, \bibinfo{pages}{139196}
  (\bibinfo{year}{2021}).

\bibitem{Tantratian2022PredictingModel}
\bibinfo{author}{Tantratian, K.}, \bibinfo{author}{Yan, H.} \&
  \bibinfo{author}{Chen, L.}
\newblock \bibinfo{title}{{Predicting pitting corrosion behavior in additive
  manufacturing: electro-chemo-mechanical phase-field model}}.
\newblock \emph{\bibinfo{journal}{Computational Materials Science}}
  \textbf{\bibinfo{volume}{213}}, \bibinfo{pages}{111640}
  (\bibinfo{year}{2022}).

\bibitem{Brewick2022SimulatingModeling}
\bibinfo{author}{Brewick, P.~T.}
\newblock \bibinfo{title}{{Simulating Pitting Corrosion in AM 316L
  Microstructures Through Phase Field Methods and Computational Modeling}}.
\newblock \emph{\bibinfo{journal}{Journal of The Electrochemical Society}}
  \textbf{\bibinfo{volume}{169}}, \bibinfo{pages}{011503}
  (\bibinfo{year}{2022}).

\bibitem{Amador2024QuantitativePhenomena}
\bibinfo{author}{Amador, J.}, \bibinfo{author}{Vega, J.},
  \bibinfo{author}{Garc{\'{i}}a-Lecina, E.} \& \bibinfo{author}{Varas, F.}
\newblock \bibinfo{title}{{Quantitative phase-field model to simulate low
  carbon steel aqueous corrosion phenomena}}.
\newblock \emph{\bibinfo{journal}{Corrosion Science}}
  \textbf{\bibinfo{volume}{232}}, \bibinfo{pages}{112045}
  (\bibinfo{year}{2024}).

\bibitem{Chadwick2018NumericalMethods}
\bibinfo{author}{Chadwick, A.~F.}, \bibinfo{author}{Stewart, J.~A.},
  \bibinfo{author}{Enrique, R.~A.}, \bibinfo{author}{Du, S.} \&
  \bibinfo{author}{Thornton, K.}
\newblock \bibinfo{title}{{Numerical Modeling of Localized Corrosion Using
  Phase-Field and Smoothed Boundary Methods}}.
\newblock \emph{\bibinfo{journal}{Journal of The Electrochemical Society}}
  \textbf{\bibinfo{volume}{165}}, \bibinfo{pages}{C633--C646}
  (\bibinfo{year}{2018}).

\bibitem{Kandekar2024MasteringScheme}
\bibinfo{author}{Kandekar, C.}, \bibinfo{author}{Ravikumar, A.},
  \bibinfo{author}{H{\"{o}}che, D.} \& \bibinfo{author}{Weber, W.~E.}
\newblock \bibinfo{title}{{Mastering the complex time-scale interaction during
  Stress Corrosion Cracking phenomena through an advanced coupling scheme}}.
\newblock \emph{\bibinfo{journal}{Computer Methods in Applied Mechanics and
  Engineering}} \textbf{\bibinfo{volume}{428}}, \bibinfo{pages}{117101}
  (\bibinfo{year}{2024}).

\bibitem{Li2023NewCorrosion}
\bibinfo{author}{Li, B.}, \bibinfo{author}{Xing, H.} \& \bibinfo{author}{Jing,
  H.}
\newblock \bibinfo{title}{{New diffusive interface model for pitting
  corrosion}}.
\newblock \emph{\bibinfo{journal}{npj Materials Degradation}}
  \textbf{\bibinfo{volume}{7}}, \bibinfo{pages}{84} (\bibinfo{year}{2023}).

\bibitem{Jafarzadeh2018PeridynamicDamage}
\bibinfo{author}{Jafarzadeh, S.}, \bibinfo{author}{Chen, Z.} \&
  \bibinfo{author}{Bobaru, F.}
\newblock \bibinfo{title}{{Peridynamic Modeling of Intergranular Corrosion
  Damage}}.
\newblock \emph{\bibinfo{journal}{Journal of The Electrochemical Society}}
  \textbf{\bibinfo{volume}{165}}, \bibinfo{pages}{C362--C374}
  (\bibinfo{year}{2018}).

\bibitem{Guiso2020IntergranularAutomata}
\bibinfo{author}{Guiso, S.}, \bibinfo{author}{di~Caprio, D.},
  \bibinfo{author}{de~Lamare, J.} \& \bibinfo{author}{Gwinner, B.}
\newblock \bibinfo{title}{{Intergranular corrosion: Comparison between
  experiments and cellular automata}}.
\newblock \emph{\bibinfo{journal}{Corrosion Science}}
  \textbf{\bibinfo{volume}{177}}, \bibinfo{pages}{108953}
  (\bibinfo{year}{2020}).

\bibitem{Guiso2022IntergranularAutomata}
\bibinfo{author}{Guiso, S.} \emph{et~al.}
\newblock \bibinfo{title}{{Intergranular corrosion in evolving media:
  Experiment and modeling by cellular automata}}.
\newblock \emph{\bibinfo{journal}{Corrosion Science}}
  \textbf{\bibinfo{volume}{205}}, \bibinfo{pages}{110457}
  (\bibinfo{year}{2022}).

\bibitem{Ansari2020Multi-Phase-FieldMaterials}
\bibinfo{author}{Ansari, T.~Q.}, \bibinfo{author}{Luo, J.-L.} \&
  \bibinfo{author}{Shi, S.-Q.}
\newblock \bibinfo{title}{{Multi-Phase-Field Model of Intergranular Corrosion
  Kinetics in Sensitized Metallic Materials}}.
\newblock \emph{\bibinfo{journal}{Journal of The Electrochemical Society}}
  \textbf{\bibinfo{volume}{167}}, \bibinfo{pages}{061508}
  (\bibinfo{year}{2020}).

\bibitem{VivekBhave2023AnSalt}
\bibinfo{author}{Vivek~Bhave, C.}, \bibinfo{author}{Zheng, G.},
  \bibinfo{author}{Sridharan, K.}, \bibinfo{author}{Schwen, D.} \&
  \bibinfo{author}{Tonks, M.~R.}
\newblock \bibinfo{title}{{An electrochemical mesoscale tool for modeling the
  corrosion of structural alloys by molten salt}}.
\newblock \emph{\bibinfo{journal}{Journal of Nuclear Materials}}
  \textbf{\bibinfo{volume}{574}}, \bibinfo{pages}{154147}
  (\bibinfo{year}{2023}).

\bibitem{Xu2008CorrosionConditions}
\bibinfo{author}{Xu, Q.} \emph{et~al.}
\newblock \bibinfo{title}{{Corrosion characteristics of low activation ferritic
  steel, JLF-1, in liquid lithium in static and thermal convection
  conditions}}.
\newblock \emph{\bibinfo{journal}{Fusion Engineering and Design}}
  \textbf{\bibinfo{volume}{83}}, \bibinfo{pages}{1477--1483}
  (\bibinfo{year}{2008}).

\bibitem{Griaznov1989InteractionConditions}
\bibinfo{author}{Griaznov, G.~M.}, \bibinfo{author}{Evtikhin, V.~A.},
  \bibinfo{author}{Zavialski, L.~P.}, \bibinfo{author}{Kosuhin, A.~Y.} \&
  \bibinfo{author}{Lyublinski, I.~E.}
\newblock \bibinfo{title}{{Interaction of structural materials with liquid
  metals under the isothermal conditions}}.
\newblock \emph{\bibinfo{journal}{Material science of liquid metal systems of
  thermonuclear reactors, Energoatomizdat}} \bibinfo{pages}{32--142}
  (\bibinfo{year}{1989}).

\bibitem{Tortorelli1984MassLithium}
\bibinfo{author}{Tortorelli, P.~F.} \& \bibinfo{author}{DeVan, J.~H.}
\newblock \bibinfo{title}{{Mass transfer behavior of a modified austenitic
  stainless steel in lithium}}.
\newblock \emph{\bibinfo{journal}{Journal of Nuclear Materials}}
  \textbf{\bibinfo{volume}{123}}, \bibinfo{pages}{1258--1263}
  (\bibinfo{year}{1984}).

\bibitem{Li2013Corrosion873K}
\bibinfo{author}{Li, Y.}, \bibinfo{author}{Abe, H.}, \bibinfo{author}{Nagasaka,
  T.}, \bibinfo{author}{Muroga, T.} \& \bibinfo{author}{Kondo, M.}
\newblock \bibinfo{title}{{Corrosion behavior of 9Cr-ODS steel in stagnant
  liquid lithium and lead–lithium at 873K}}.
\newblock \emph{\bibinfo{journal}{Journal of Nuclear Materials}}
  \textbf{\bibinfo{volume}{443}}, \bibinfo{pages}{200--206}
  (\bibinfo{year}{2013}).

\bibitem{Hariharan2024MicrostructureDesign}
\bibinfo{author}{Hariharan, K.} \& \bibinfo{author}{Virtanen, S.}
\newblock \bibinfo{title}{{Microstructure engineering for corrosion resistance
  in structural alloy design}}.
\newblock \emph{\bibinfo{journal}{npj Materials Degradation}}
  \textbf{\bibinfo{volume}{8}}, \bibinfo{pages}{115} (\bibinfo{year}{2024}).

\bibitem{DiStefano1964CorrosionLithium}
\bibinfo{author}{DiStefano, J.}
\newblock \emph{\bibinfo{title}{{Corrosion of Refractory Metals by Lithium}}}.
\newblock Ph.D. thesis, \bibinfo{school}{Oak Ridge National Laboratory (ORNL)},
  \bibinfo{address}{Oak Ridge, TN (United States)} (\bibinfo{year}{1964}).

\bibitem{Kondo2010Erosion-corrosionImpeller}
\bibinfo{author}{Kondo, M.}, \bibinfo{author}{Muroga, T.},
  \bibinfo{author}{Nagasaka, T.}, \bibinfo{author}{Tsisar, V.} \&
  \bibinfo{author}{Yeliseyeva, O.}
\newblock \bibinfo{title}{{Erosion-corrosion of RAFM JLF-1 steel in lithium
  flow induced by impeller}}.
\newblock \emph{\bibinfo{journal}{Journal of Plasma and Fusion Research}}
  \textbf{\bibinfo{volume}{9}}, \bibinfo{pages}{294--299}
  (\bibinfo{year}{2010}).

\bibitem{Kondo2011FlowBreeders}
\bibinfo{author}{Kondo, M.} \emph{et~al.}
\newblock \bibinfo{title}{{Flow accelerated corrosion and erosion–corrosion
  of RAFM steel in liquid breeders}}.
\newblock \emph{\bibinfo{journal}{Fusion Engineering and Design}}
  \textbf{\bibinfo{volume}{86}}, \bibinfo{pages}{2500--2503}
  (\bibinfo{year}{2011}).

\bibitem{Tsisar2011EffectLithium}
\bibinfo{author}{Tsisar, V.} \emph{et~al.}
\newblock \bibinfo{title}{{Effect of nitrogen on the corrosion behavior of RAFM
  JLF-1 steel in lithium}}.
\newblock \emph{\bibinfo{journal}{Journal of Nuclear Materials}}
  \textbf{\bibinfo{volume}{417}}, \bibinfo{pages}{1205--1209}
  (\bibinfo{year}{2011}).

\bibitem{Beskorovaynyi1983MechanismsLithium}
\bibinfo{author}{Beskorovaynyi, N.~M.} \& \bibinfo{author}{Ioltuhovski, A.~G.}
\newblock \bibinfo{title}{{Mechanisms of corrosion processes under the contact
  of structural materials with liquid lithium}}.
\newblock \emph{\bibinfo{journal}{Structural Materials and Liquid Metal
  Heat-transfers, Energoatomizdat, Moscow}} \bibinfo{pages}{59--89}
  (\bibinfo{year}{1983}).

\bibitem{Lee2005EffectSteel}
\bibinfo{author}{Lee, S.~J.} \& \bibinfo{author}{Lee, Y.-K.}
\newblock \bibinfo{title}{{Effect of Austenite Grain Size on Martensitic
  Transformation of a Low Alloy Steel}}.
\newblock \emph{\bibinfo{journal}{Materials Science Forum}}
  \textbf{\bibinfo{volume}{475-479}}, \bibinfo{pages}{3169--3172}
  (\bibinfo{year}{2005}).

\bibitem{Souza2020AustenitizingMicroalloyed-Steel}
\bibinfo{author}{Souza, S. d. S.~d.}, \bibinfo{author}{Moreira, P.~S.} \&
  \bibinfo{author}{Faria, G. L.~d.}
\newblock \bibinfo{title}{{Austenitizing Temperature and Cooling Rate Effects
  on the Martensitic Transformation in a Microalloyed-Steel}}.
\newblock \emph{\bibinfo{journal}{Materials Research}}
  \textbf{\bibinfo{volume}{23}} (\bibinfo{year}{2020}).

\bibitem{Tortorelli1981CompatibilityLithium}
\bibinfo{author}{Tortorelli, P.~F.}, \bibinfo{author}{Devan, J.~H.} \&
  \bibinfo{author}{Yonco, R.~M.}
\newblock \bibinfo{title}{{Compatibility of Fe-Cr-Mo alloys with static
  lithium}}.
\newblock \emph{\bibinfo{journal}{Journal of Materials for Energy Systems}}
  \textbf{\bibinfo{volume}{2}}, \bibinfo{pages}{5--15} (\bibinfo{year}{1981}).

\bibitem{Zhang2015SimulatingStructure}
\bibinfo{author}{Zhang, J.}, \bibinfo{author}{Chen, Z.~H.} \&
  \bibinfo{author}{Dong, C.~F.}
\newblock \bibinfo{title}{{Simulating Intergranular Stress Corrosion Cracking
  in AZ31 Using Three-Dimensional Cohesive Elements for Grain Structure}}.
\newblock \emph{\bibinfo{journal}{Journal of Materials Engineering and
  Performance}} \textbf{\bibinfo{volume}{24}}, \bibinfo{pages}{4908--4918}
  (\bibinfo{year}{2015}).

\bibitem{Nguyen2017Multi-phase-fieldMaterials}
\bibinfo{author}{Nguyen, T.-T.}, \bibinfo{author}{R{\'{e}}thor{\'{e}}, J.},
  \bibinfo{author}{Yvonnet, J.} \& \bibinfo{author}{Baietto, M.-C.}
\newblock \bibinfo{title}{{Multi-phase-field modeling of anisotropic crack
  propagation for polycrystalline materials}}.
\newblock \emph{\bibinfo{journal}{Computational Mechanics}}
  \textbf{\bibinfo{volume}{60}}, \bibinfo{pages}{289--314}
  (\bibinfo{year}{2017}).

\bibitem{Simonovski2011ComputationalCracking}
\bibinfo{author}{Simonovski, I.} \& \bibinfo{author}{Cizelj, L.}
\newblock \bibinfo{title}{{Computational multiscale modeling of intergranular
  cracking}}.
\newblock \emph{\bibinfo{journal}{Journal of Nuclear Materials}}
  \textbf{\bibinfo{volume}{414}}, \bibinfo{pages}{243--250}
  (\bibinfo{year}{2011}).

\bibitem{Reeves1976GrainSteel}
\bibinfo{author}{Reeves, J.~A.}, \bibinfo{author}{Olson, D.~L.} \&
  \bibinfo{author}{Bradley, W.~L.}
\newblock \bibinfo{title}{{Grain Boundary Penetration Kinetics of Nitrided Type
  304L Stainless Steel}}.
\newblock \emph{\bibinfo{journal}{Nuclear Technology}}
  \textbf{\bibinfo{volume}{30}}, \bibinfo{pages}{385--389}
  (\bibinfo{year}{1976}).

\bibitem{Patterson1975LithiumSteel}
\bibinfo{author}{Patterson, R.}, \bibinfo{author}{Schlager, R.} \&
  \bibinfo{author}{Olson, D.}
\newblock \bibinfo{title}{{Lithium grain-boundary penetration of 304L stainless
  steel}}.
\newblock \emph{\bibinfo{journal}{Journal of Nuclear Materials}}
  \textbf{\bibinfo{volume}{57}}, \bibinfo{pages}{312--316}
  (\bibinfo{year}{1975}).

\bibitem{Wheeler1992Phase-fieldAlloys}
\bibinfo{author}{Wheeler, A.~A.}, \bibinfo{author}{Boettinger, W.~J.} \&
  \bibinfo{author}{McFadden, G.~B.}
\newblock \bibinfo{title}{{Phase-field model for isothermal phase transitions
  in binary alloys}}.
\newblock \emph{\bibinfo{journal}{Physical Review A}}
  \textbf{\bibinfo{volume}{45}}, \bibinfo{pages}{7424--7439}
  (\bibinfo{year}{1992}).

\bibitem{Hu2007ThermodynamicApproach}
\bibinfo{author}{Hu, S.}, \bibinfo{author}{Murray, J.},
  \bibinfo{author}{Weiland, H.}, \bibinfo{author}{Liu, Z.} \&
  \bibinfo{author}{Chen, L.}
\newblock \bibinfo{title}{{Thermodynamic description and growth kinetics of
  stoichiometric precipitates in the phase-field approach}}.
\newblock \emph{\bibinfo{journal}{Calphad}} \textbf{\bibinfo{volume}{31}},
  \bibinfo{pages}{303--312} (\bibinfo{year}{2007}).

\bibitem{Kim1999Phase-fieldAlloys}
\bibinfo{author}{Kim, S.~G.}, \bibinfo{author}{Kim, W.~T.} \&
  \bibinfo{author}{Suzuki, T.}
\newblock \bibinfo{title}{{Phase-field model for binary alloys}}.
\newblock \emph{\bibinfo{journal}{Physical Review E}}
  \textbf{\bibinfo{volume}{60}}, \bibinfo{pages}{7186--7197}
  (\bibinfo{year}{1999}).

\bibitem{Makuch2024ALayer}
\bibinfo{author}{Makuch, M.}, \bibinfo{author}{Kovacevic, S.},
  \bibinfo{author}{Wenman, M.~R.} \&
  \bibinfo{author}{Mart{\'{i}}nez-Pa{\~{n}}eda, E.}
\newblock \bibinfo{title}{{A nonlinear phase-field model of corrosion with
  charging kinetics of electric double layer}}.
\newblock \emph{\bibinfo{journal}{Modelling and Simulation in Materials Science
  and Engineering}} \textbf{\bibinfo{volume}{32}}, \bibinfo{pages}{075012}
  (\bibinfo{year}{2024}).

\bibitem{Kovacevic2020InterfacialCeramics}
\bibinfo{author}{Kovacevic, S.}, \bibinfo{author}{Pan, R.},
  \bibinfo{author}{Sekulic, D.} \& \bibinfo{author}{Mesarovic, S.}
\newblock \bibinfo{title}{{Interfacial energy as the driving force for
  diffusion bonding of ceramics}}.
\newblock \emph{\bibinfo{journal}{Acta Materialia}}
  \textbf{\bibinfo{volume}{186}}, \bibinfo{pages}{405--414}
  (\bibinfo{year}{2020}).

\bibitem{Kovacevic2023Phase-fieldApplications}
\bibinfo{author}{Kovacevic, S.}, \bibinfo{author}{Ali, W.},
  \bibinfo{author}{Mart{\'{i}}nez-Pa{\~{n}}eda, E.} \& \bibinfo{author}{LLorca,
  J.}
\newblock \bibinfo{title}{{Phase-field modeling of pitting and
  mechanically-assisted corrosion of Mg alloys for biomedical applications}}.
\newblock \emph{\bibinfo{journal}{Acta Biomaterialia}}
  \textbf{\bibinfo{volume}{164}}, \bibinfo{pages}{641--658}
  (\bibinfo{year}{2023}).

\bibitem{Allen1979ACoarsening}
\bibinfo{author}{Allen, S.~M.} \& \bibinfo{author}{Cahn, J.~W.}
\newblock \bibinfo{title}{{A microscopic theory for antiphase boundary motion
  and its application to antiphase domain coarsening}}.
\newblock \emph{\bibinfo{journal}{Acta Metallurgica}}
  \textbf{\bibinfo{volume}{27}}, \bibinfo{pages}{1085--1095}
  (\bibinfo{year}{1979}).

\bibitem{Mai2018NewProcesses}
\bibinfo{author}{Mai, W.} \& \bibinfo{author}{Soghrati, S.}
\newblock \bibinfo{title}{{New phase field model for simulating galvanic and
  pitting corrosion processes}}.
\newblock \emph{\bibinfo{journal}{Electrochimica Acta}}
  \textbf{\bibinfo{volume}{260}}, \bibinfo{pages}{290--304}
  (\bibinfo{year}{2018}).

\bibitem{Simon2022MechanisticParticles}
\bibinfo{author}{Simon, P.-C.}, \bibinfo{author}{Aagesen, L.~K.},
  \bibinfo{author}{Jiang, C.}, \bibinfo{author}{Jiang, W.} \&
  \bibinfo{author}{Ke, J.-H.}
\newblock \bibinfo{title}{{Mechanistic calculation of the effective silver
  diffusion coefficient in polycrystalline silicon carbide: Application to
  silver release in AGR-1 TRISO particles}}.
\newblock \emph{\bibinfo{journal}{Journal of Nuclear Materials}}
  \textbf{\bibinfo{volume}{563}}, \bibinfo{pages}{153669}
  (\bibinfo{year}{2022}).

\bibitem{Nguyen2016AMicrotomography}
\bibinfo{author}{Nguyen, T.}, \bibinfo{author}{Yvonnet, J.},
  \bibinfo{author}{Zhu, Q.-Z.}, \bibinfo{author}{Bornert, M.} \&
  \bibinfo{author}{Chateau, C.}
\newblock \bibinfo{title}{{A phase-field method for computational modeling of
  interfacial damage interacting with crack propagation in realistic
  microstructures obtained by microtomography}}.
\newblock \emph{\bibinfo{journal}{Computer Methods in Applied Mechanics and
  Engineering}} \textbf{\bibinfo{volume}{312}}, \bibinfo{pages}{567--595}
  (\bibinfo{year}{2016}).

\bibitem{COMSOLSweden}
\bibinfo{title}{{COMSOL Multiphysics v. 6.0. https://www.comsol.com. COMSOL AB,
  Stockholm, Sweden}} .

\bibitem{Coen1984CompatibilityIspra}
\bibinfo{author}{Coen, V.} \& \bibinfo{author}{Fenici, P.}
\newblock \bibinfo{title}{{Compatibility of structural materials with liquid
  breeders — A review of recent work carried out at JRC, Ispra}}.
\newblock \emph{\bibinfo{journal}{Nuclear Engineering and Design. Fusion}}
  \textbf{\bibinfo{volume}{1}}, \bibinfo{pages}{215--229}
  (\bibinfo{year}{1984}).

\bibitem{Knaster2017AssessmentStudy}
\bibinfo{author}{Knaster, J.} \& \bibinfo{author}{Favuzza, P.}
\newblock \bibinfo{title}{{Assessment of corrosion phenomena in liquid lithium
  at 873 K. A Li(d,n) neutron source as case study}}.
\newblock \emph{\bibinfo{journal}{Fusion Engineering and Design}}
  \textbf{\bibinfo{volume}{118}}, \bibinfo{pages}{135--141}
  (\bibinfo{year}{2017}).

\bibitem{Ruedl1983EffectSteel}
\bibinfo{author}{Ruedl, E.} \& \bibinfo{author}{Sasaki, T.}
\newblock \bibinfo{title}{{Effect of lithium on grain-boundary precipitation in
  a Cr-Mn austenitic steel}}.
\newblock \emph{\bibinfo{journal}{Journal of Nuclear Materials}}
  \textbf{\bibinfo{volume}{116}}, \bibinfo{pages}{112--122}
  (\bibinfo{year}{1983}).

\bibitem{Cermak1996Low-temperatureAlloys}
\bibinfo{author}{{\v{C}}erm{\'{a}}k, J.},
  \bibinfo{author}{Rů{\v{z}}i{\v{c}}kov{\'{a}}, J.} \&
  \bibinfo{author}{Pokorn{\'{a}}, A.}
\newblock \bibinfo{title}{{Low-temperature tracer diffusion of chromium in
  Fe-Cr ferritic alloys}}.
\newblock \emph{\bibinfo{journal}{Scripta Materialia}}
  \textbf{\bibinfo{volume}{35}}, \bibinfo{pages}{411--416}
  (\bibinfo{year}{1996}).

\bibitem{Li2023TheoreticalMetals}
\bibinfo{author}{Li, C.}, \bibinfo{author}{Lu, S.}, \bibinfo{author}{Divinski,
  S.} \& \bibinfo{author}{Vitos, L.}
\newblock \bibinfo{title}{{Theoretical and experimental grain boundary energies
  in body-centered cubic metals}}.
\newblock \emph{\bibinfo{journal}{Acta Materialia}}
  \textbf{\bibinfo{volume}{255}}, \bibinfo{pages}{119074}
  (\bibinfo{year}{2023}).

\bibitem{Scheiber2016AbMetals}
\bibinfo{author}{Scheiber, D.}, \bibinfo{author}{Pippan, R.},
  \bibinfo{author}{Puschnig, P.} \& \bibinfo{author}{Romaner, L.}
\newblock \bibinfo{title}{{Ab initio calculations of grain boundaries in bcc
  metals}}.
\newblock \emph{\bibinfo{journal}{Modelling and Simulation in Materials Science
  and Engineering}} \textbf{\bibinfo{volume}{24}}, \bibinfo{pages}{035013}
  (\bibinfo{year}{2016}).

\bibitem{Nakamichi2008QuantitativeFE-STEM}
\bibinfo{author}{Nakamichi, H.}, \bibinfo{author}{Sato, K.},
  \bibinfo{author}{Miyata, Y.}, \bibinfo{author}{Kimura, M.} \&
  \bibinfo{author}{Masamura, K.}
\newblock \bibinfo{title}{{Quantitative analysis of Cr-depleted zone morphology
  in low carbon martensitic stainless steel using FE-(S)TEM}}.
\newblock \emph{\bibinfo{journal}{Corrosion Science}}
  \textbf{\bibinfo{volume}{50}}, \bibinfo{pages}{309--315}
  (\bibinfo{year}{2008}).

\bibitem{KumarThakur2023ASystems}
\bibinfo{author}{Kumar~Thakur, A.}, \bibinfo{author}{Kovacevic, S.},
  \bibinfo{author}{Manga, V.~R.}, \bibinfo{author}{Deymier, P.~A.} \&
  \bibinfo{author}{Muralidharan, K.}
\newblock \bibinfo{title}{{A first-principles and CALPHAD-assisted phase-field
  model for microstructure evolution: Application to Mo-V binary alloy
  systems}}.
\newblock \emph{\bibinfo{journal}{Materials {\&} Design}}
  \textbf{\bibinfo{volume}{235}}, \bibinfo{pages}{112443}
  (\bibinfo{year}{2023}).

\end{thebibliography}
\end{document}